\documentclass[acmsmall]{acmart}

\usepackage{booktabs}
\usepackage{tikz}
\usetikzlibrary{external}
\usepackage{ textcomp }
\usepackage[utf8]{inputenc}
\usepackage{enumitem}

\usepackage{epstopdf}
\usepackage{makecell}
\usepackage{threeparttable}
\usepackage{longtable}
\usepackage{multirow}
\usepackage{tabularx}
\usepackage{tabulary}
\usepackage{tablefootnote}
\usepackage{array}
\newcommand{\PreserveBackslash}[1]{\let\temp=\\#1\let\\=\temp}

\newcolumntype{C}[1]{>{\PreserveBackslash\centering}p{#1}}
\newcolumntype{R}[1]{>{\PreserveBackslash\raggedleft}p{#1}}
\newcolumntype{L}[1]{>{\PreserveBackslash\raggedright}p{#1}}
\usepackage{dsfont}
\usepackage{epstopdf}
\usepackage{subfigure}
\usepackage{graphicx}
\usepackage{caption}
\usepackage{color}
\usepackage{mathtools}
\usepackage{url}
\usepackage{hyperref}
\usepackage{bm}
\usepackage{graphicx}
\usepackage{algorithm}
\usepackage{algpseudocode}
\usepackage{amsmath}

\usepackage{amsfonts}
\usepackage{subfigure}
\usepackage{multirow}
\usepackage{bm}
\usepackage{graphicx}
\usepackage{booktabs}
\usepackage{wrapfig}

\acmArticle{101}

\newcommand{\bruce}[1]{\textcolor{black}{#1}}
\newcommand{\wan}[1]{\textcolor{black}{#1}}

\newcommand{\xin}[1]{\textcolor{black}{#1}}  %
\newcommand{\red}[1]{\textcolor{black}{#1}}
\newcommand{\newblue}[1]{\textcolor{blue}{#1}}
\begin{document}

\title{Spatio-Temporal Contrastive Learning Enhanced GNNs for Session-based Recommendation}

\author{Zhongwei Wan$^{\dagger}$}
\thanks{$^{\dag}$ denotes currently affiliated with The Ohio State University.}
\authornote{The work has been accepted by ACM Transaction on Information Systems (ACM TOIS).}
\affiliation{
  \institution{The Ohio State University}
  \country{USA}
}
\email{wan.512@osu.edu}

\author{Xin Liu}
\affiliation{%
  \institution{Hong Kong University of Science and Technology}
  \country{China}
}
\email{xliucr@cse.ust.hk}

\author{Benyou Wang}
\affiliation{%
  \institution{The Chinese University of Hong Kong, Shenzhen}
  \country{China}
}
\email{wangbenyou@cuhk.edu.cn}

\author{Jiezhong Qiu}
\affiliation{%
  \institution{Tencent}
  \country{China}
}
\email{xptree@foxmail.com}

\author{Boyu Li}
\affiliation{%
  \institution{University of Technology Sydney}
  \country{Australia}
}
\email{Boyu.Li@student.uts.edu.au}

\author{Ting Guo}
\affiliation{%
  \institution{University of Technology Sydney}
  \country{Australia}
}
\email{Ting.Guo@uts.edu.au}

\author{Guangyong Chen}
\authornote{Corresponding author.}
\affiliation{%
  \institution{Zhejiang Lab}
  \country{China}
}
\email{gychen@zhejianglab.com}

\author{Yang Wang}
\affiliation{%
  \institution{University of Technology Sydney}
  \country{Australia}
}
\email{Yang.Wang@uts.edu.au}
\begin{abstract}
Session-based recommendation (SBR) systems aim to utilize the user's short-term behavior sequence to predict the next item without the detailed user profile.
\wan{Most recent works try to model the user preference by treating the sessions as between-item transition graphs and utilize various graph neural networks (GNNs) to encode the representations of pair-wise relations among items and their neighbors. Some of the existing GNN-based models mainly focus on aggregating information from the view of spatial graph structure, which ignores the temporal relations within neighbors of an item during message passing and the information loss results in a sub-optimal problem. Other works embrace this challenge by incorporating additional temporal information but lack sufficient interaction between the spatial and temporal patterns. To address this issue, inspired by the uniformity and alignment properties of contrastive learning techniques, we propose a novel framework called Session-based \textbf{RE}commendation with \textbf{S}patio-\textbf{T}emporal \textbf{C}ontrastive Learning Enhanced GNNs (\textbf{RESTC}). The idea is to supplement the GNN-based main supervised recommendation task with the temporal representation via an auxiliary cross-view contrastive learning mechanism.
Furthermore, a novel global collaborative filtering graph (CFG) embedding is leveraged to enhance the spatial view in the main task.} Extensive experiments demonstrate the significant performance of RESTC compared with the state-of-the-art baselines. We release our source code at \newblue{https://github.com/SUSTechBruce/RESTC-Source-code}.

\end{abstract}


\maketitle

\section{Introduction}
Recommendation \wan{systems have been an efficient tool for helping users make informative choices according to their available profiles and the preferences reflected in the long-term history \xin{interactions}, which are widely used in web search and various stream medias~\cite{DBLP:conf/www/HeLZNHC17, DBLP:journals/csur/ZhangYST19, DBLP:journals/csur/WangCWSOL22}. However, the traditional recommenders may perform poorly in some scenarios where the user's \xin{interactions} are \xin{inadequate} in a narrow period, or the status is unlogged-in. Thus, Session-based Recommendation (SBR) has attracted increasing research \cite{DBLP:conf/cikm/LiRCRLM17, DBLP:conf/aaai/WuT0WXT19, DBLP:conf/ijcai/XuZLSXZFZ19, DBLP:journals/tii/ChenHWL22}, since it characterizes users' short-term preference from the limited interactions in the current session, e.g., a basket of products purchased in one transaction visit, and then predict the products that a user interacts with in the future.}

\begin{figure}[htbp]
  \centering
  \includegraphics[width=0.75\linewidth]{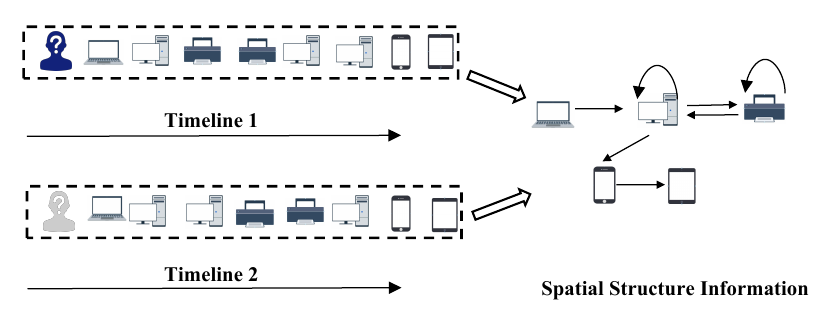}
  \vspace{-0.1in}
  \caption{Two distinct sessions may be represented as the same graph if the temporal information is omitted, indicating the temporal pattern should be sufficiently considered to supplement GNN-based models for SBR task. }
  \label{fig:motivation2}
  \vspace{-0.1in}
\end{figure}

\begin{figure*}[t]
  \centering
  \includegraphics[width=1\linewidth]{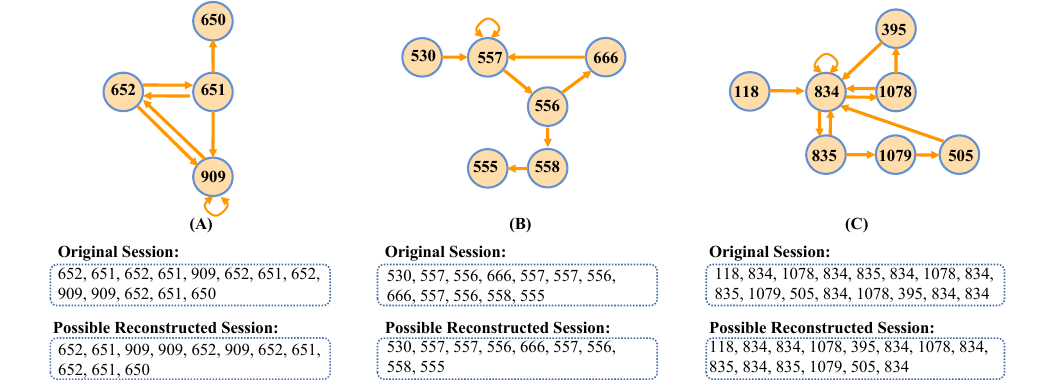}
  \vspace{-0.2in}
  \caption {Sampling sessions from real public dataset. The numbers in the nodes denotes the index of items.}
  \vspace{-0.2in}
  \label{fig:sample real data}
\end{figure*}

\wan{Recently, most existing SBR methods~\cite{DBLP:conf/aaai/WuT0WXT19,DBLP:conf/cikm/QiuLHY19,DBLP:conf/ijcai/XuZLSXZFZ19,DBLP:conf/ijcai/LuoZLZWXFS20,DBLP:conf/kdd/ChenW20} mainly construct the graph structure from the session and leverage Graph Neural Networks (GNNs) to conduct information aggregation between adjacent items and capture complex high-order relations, which have obtained effective performance.}
However, the temporal information has been omitted by the abovementioned GNN-based methods because of the permutation-invariant aggregation during the message passing in the graph structure, which is a vital signal that contributes significantly to capturing the preference evolution of the user in the temporal dimension~\cite{DBLP:conf/kdd/KumarZL19, DBLP:conf/cikm/FanLZX0Y21}. Fig~\ref{fig:motivation2} shows a concrete example of the temporal information loss's impact. Suppose the two sessions produce the different next item, but they are encoded as the same graph representation since the aggregation function of GNN could not distinguish the temporal order of items' neighbors. In that case, the GNN-based model will induce incorrect results and limit its capacity without the essential temporal pattern. 
\bruce{Additionally, to better understand the importance of temporal information among session graphs, we randomly sample some representative session data from the public dataset Dignetica~\footnote{http://2015.recsyschallenge.com/challenge.html}. As shown in Fig~\ref{fig:sample real data}, to process sessions using GNNs, the original sessions need to be converted into graphs first. However, from (A), (B) and (C) of Fig~\ref{fig:sample real data}, we notice that it is difficult to accurately reconstruct the original session if these session graphs contain directed cycles.} 
\red{More specifically, the showcases of directed rings are notable examples of samples with in-degree exceeding 1. We count the presence of directed graphs in the Diginetica~\footnote{http://2015.recsyschallenge.com/challenge.ht} dataset with in-degree greater than 1. The sequences in which the in-degree is greater than 1 account for 23.85$\%$ of the overall samples.
Therefore, those phenomenons also indicate the necessity of incorporating temporal patterns.}

Fortunately, some  works have attempted to incorporate temporal information by modeling a session as the dynamic sub-session graphs at the fixed-length time intervals~\cite{Pan2021DynamicGL,DBLP:conf/sigir/ZhouTHZW21} or integrating the timestamps information as a contextual dimension~\cite{Shen2021TemporalAM}. However, modeling multiple sub-session graphs based on \xin{timelines} may introduce redundant spatial structure information and it still misses temporal \xin{orders} during aggregation. 
The other lines of works directly treat each session as a sequence of items with the relative order or position information and utilize Recurrent Neural Networks (RNN)~\cite{DBLP:conf/recsys/TanXL16, DBLP:conf/cikm/HidasiK18, DBLP:journals/corr/HidasiKBT15} or memory networks~\cite{DBLP:conf/kdd/LiuZMZ18} to learn the sequential signal in a session to capture users' preference. But modeling sequences using RNN-based models are arguably insufficient to obtain accurate user representation in sessions and neglect complex transition patterns of items~\cite{DBLP:conf/aaai/WuT0WXT19}. 
Besides, all of these methods lack sufficient \xin{interactions} between spatial \xin{structures} and temporal patterns in the latent space, which restricts the representation capability of the models.
\vspace{-0.3mm}

Therefore, incorporating temporal pattern then modeling the latent mutual presentation of spatial and temporal views of a session is crucial and challenging for session-based recommendation systems. To \wan{align the embeddings of the two views in a unified latent space},
(i) one straightforward way could be to directly adopt concatenation or cross-attention based methods~\cite{DBLP:conf/nips/LiSGJXH21, DBLP:journals/corr/abs-2111-08276} to fuse these two information resources after the encoding phase. But both views know little information about each other in this way since there is no efficient \xin{interactions} between two different encoders during training\xin{;} (ii) \xin{the} other approaches could be to utilize GAN \cite{DBLP:conf/nips/LiuT16, liang2020large, liang2020many} to learn the joint distribution of multi-style views or leverage semi-supervised learning \xin{paradigms} like Co-training \cite{DBLP:conf/colt/BlumM98} to acquire complementary information from each other views.
However, it is unstable to optimize the min-max objective of GAN-style methods. Besides, both GAN and Co-training mechanisms face the mode collapse problem~\cite{DBLP:conf/eccv/QiaoSZWY18} while learning the latent \xin{representations} of different views during training. 

Due to the issues mentioned above, inspired by the uniformity property~\cite{DBLP:conf/wsdm/QiuHYW22} and theoretical guarantee of semantic representation alignment in latent space~\cite{DBLP:conf/icml/SaunshiPAKK19} for contrastive learning, we propose a novel auxiliary \textbf{spatio-temporal contrastive learning} framework named RESTC. RESTC can align the spatial and temporal semantic representations in a projected feature space to conserve as much mutual information of the two views as possible. Although existing contrastive learning techniques for sequential~\cite{https://doi.org/10.48550/arxiv.2010.14395, DBLP:journals/corr/abs-2108-06479, DBLP:conf/www/ChenLLMX22} or GNN-based recommendation~\cite{DBLP:journals/corr/abs-2111-08268, DBLP:conf/aaai/HuangCXXDCBZH21, DBLP:conf/sigir/WuWF0CLX21} generally generate positive samples using item-level augmentation, e.g., item cropping, masking, reordering or \xin{sub-sampling} in sequence and graph data, respectively, which are not suitable for SBR task since these methods induce semantically inconsistent samples and damage the completeness of temporal \xin{patterns}. \bruce{Specifically, sequential augmentation (e.g., cropping, masking, re-ordering) may compromise the completeness of users' original intent and their evolutionary preference~\cite{Guo2022EvolutionaryPL} related to temporal order, which may introduce more bias during training. Therefore, completely capturing the temporal pattern is essential for modelling the user's interest. Meanwhile, it avoids additional noise caused by damaging the temporal order. Furthermore, since the average number of nodes of session-based graphs shown in Tab~\ref{tab:data} is relatively small compared with user-item interacted graphs~\cite{DBLP:conf/sigir/WuWF0CLX21}, the graph augmentation (e.g., sub-sampling) is hard to sample useful sub-structural information and also disrupt the completeness of spatial information, which may also introduce bias during sub-graph contrastive learning.}

\begin{figure*}[h]
  \centering
  \includegraphics[width=1\linewidth]{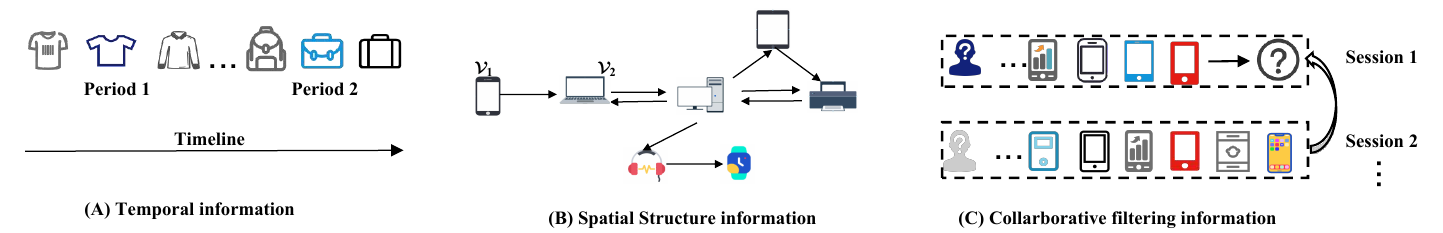}
  \vspace{-0.2in}
  \caption {Three essential information among sessions data:
  (A) \textit{temporal view of a session} is about a behavioral sequence containing user's dynamic preference w.r.t its timeline;
  (B) \textit{spatial view of a session} refers to a between-item transition directed graph, each edge of which indicates a behavior shift from the source item to the target item --- for example, a user has clicked item $v_2$ after $v_1$. Note that behavior shift associated with an edge could happen many times in a session, and such edges are orthogonal to time;
  (C) \textit{collaborative filtering information in other sessions} could be extracted from a global weighted graph then used to compensate for the item profiles in a short-term session. 
   }
  \vspace{-0.2in}
  \label{fig:motivation}
\end{figure*}

Different from the above works, we comprehensively consider two views on the session level and adopt a spatial encoder for the graph structure representation and a temporal encoder to supplement the temporal representation as the informative, positive sample.
\wan{Specifically, it is worthwhile to notice that our RESTC is model-agnostic that can be applied to any GNN-based model. Here we employ the powerful Multi-relational Graph Attention Network (MGAT) refined by GAT ~\cite{DBLP:conf/iclr/VelickovicCCRLB18} as the spatial encoder. We further derive a well-designed Session Transformer (SESTrans) augmented with a temporal enhanced module as the temporal encoder. For the contrastive objective, we propose a mixed noise negative sampling strategy different from~\cite{DBLP:conf/icml/ChenK0H20} to further enhance the model performance. With the contrastive learning loss, we enhance the cross-view \xin{interactions} in the latent space to refine session representation by maximizing the agreement of positive pairs. Furthermore, due to the data sparsity of short-term session data, a Collaborative Filtering Graph (CFG) derived from all sessions as a global weighted item transition graph, is leveraged to enhance the spatial view with the collaborative filtering embedding in the main supervised task. The example of spatial, temporal  and collaborative information of a session are shown in Fig~\ref{fig:motivation}, and the pipeline of RESTC is illustrated in Fig~\ref{fig:architecture}. The detailed experiment results show that RESTC outperforms the state-of-the-art baselines, showing the effectiveness of incorporating temporal information with spatio-temporal contrastive learning.} Our \textbf{contributions} can be summarized \xin{below}.

\begin{itemize}[leftmargin=*]

\setlength{\itemsep}{1.0pt}
   \setlength{\parsep}{1.0pt}
   \setlength{\parskip}{1.0pt}
    \item \wan{We highlight the significance of incorporating temporal information for GNN-based SBR task, facilitating the development of cross-view \xin{interactions} for the spatial and temporal pattern.}
    
    \item \wan{To the best of our knowledge, the proposed spatio-temporal contrastive learning framework RESTC is the first work aiming to align and refine the representations of spatial and temporal views in the latent space for the SBR task, which can effectively plugged into many existing GNN-based models. }
    
     \item \wan{We conduct extensive experiments on six real-life public datasets, demonstrating that our model consistently outperforms the state-of-the-art methods with a large margin.}
\end{itemize}

\section{Related Work}
\label{sec:paradigms}

\subsection{Sequence-based Models in SBR}

In early research, FPMC \cite{DBLP:conf/www/RendleFS10} utilized Markov chain and matrix factorization to obtain the sequential pattern of session. Recently,  neural network-based models have demonstrated effectiveness in exploiting sequential data in SBR tasks. GRU4Rec \cite{DBLP:journals/corr/HidasiKBT15} was the first RNN-based model which captured item transitions by multi-layer GRUs. NARM \cite{DBLP:conf/cikm/LiRCRLM17} leveraged an attention-based method to combine RNN to model complex item relations better. STAMP \cite{DBLP:conf/kdd/LiuZMZ18}  used the attention-based memory network to capture the user's current interest.
Inspired by Transformer architecture, SASRec \cite{Kang2018SelfAttentiveSR} stacked several self-attention layers to model the item-transition sequence. BERT4Rec \cite{DBLP:conf/cikm/SunLWPLOJ19} employed deep bidirectional self-attention to model user behaviors for sequence recommendation. Besides, Yuan et al. \cite{DBLP:conf/aaai/YuanSSWZ21} also propose to use a dual sparse attention network to explore the current user's interest via an adaptively learnable target embedding. These attention-based models separately deal with the user’s last item and the whole current session, thus capturing the user’s general and recent interest. 


\subsection{GNN-based Models in SBR}
\wan{ Most recent works focus on utilizing Graph Neural Networks (GNNs) to extract the relationship in the session, which have shown better results than sequence-based models~\cite{DBLP:conf/aaai/WuT0WXT19, DBLP:conf/ijcai/XuZLSXZFZ19, DBLP:conf/cikm/QiuLHY19}.}
For instance, SR-GNN \cite{DBLP:conf/aaai/WuT0WXT19} used a gate GNN model to obtain item embeddings over an item graph and predict the next item using the attention mechanism. GC-SAN \cite{DBLP:conf/ijcai/XuZLSXZFZ19} utilized self-attention networks to aggregate the information of session graphs. FGNN \cite{DBLP:conf/cikm/QiuLHY19} leveraged multi-head attention to aggregate the neighbor item's embeddings in a weighted item-transition graph. LESSR~\cite{DBLP:conf/kdd/ChenW20}
preserved session order  based on GRU and shortcut graph attention to solve the lossy session encoding and ineffective long-range dependency capturing problems. 
\wan{Zhou and Pan et al.~\cite{DBLP:conf/sigir/ZhouTHZW21,Pan2021DynamicGL} constructed a sequence of dynamic graph snapshots at timestamps to model the preference evolution.} GCE-GNN ~\cite{DBLP:conf/sigir/Wang0CLMQ20} proposed to exploit a session-graph convolution and global neighbor graph convolution to conduct a more accurate session embedding. GCARM~\cite{DBLP:journals/tois/PanCCC22} considered the dynamic correlations between the local and global neighbors of each node during the information propagation.
\bruce{G$^{3}$-SR~\cite{Deng2022G3SRGG} proposed global graph
guided SBR by leveraging an unsupervised pre-training process to extract global item-to-item relational information.
However, traditional GNN-based models lack temporal information to capture users' evolutionary preferences.}

\subsection{Temporal Augmented Models in RS}
\bruce{Temporal information can model users' dynamic preferences over time and play an essential role in the recommendation system. Some previous works have incorporated temporal patterns into GNNs in other recommendation settings~\cite{Kumar2019PredictingDE, Xu2020InductiveRL, Wang2021AdaptiveDA, Shen2021TemporalAM, Zhang2021DynamicGN}. For instance, JODIE~\cite{Kumar2019PredictingDE} designed a coupled recurrent neural network model that learns users' embedding trajectories and estimates the user's embedding at any time in the future. 
TGAT~\cite{Xu2020InductiveRL} proposed the temporal graph attention layer to aggregate temporal-topological neighbourhood features and learn the time-feature interaction efficiently.
TGN-MetA~\cite{Shen2021TemporalAM} utilized the memory-tower augmentation to process the augmented graphs of different magnitudes on separate levels to optimize Temporal GCNs.
DGSR~\cite{Zhang2021DynamicGN} proposed to explore the interactive behaviour of users and items with time and order information by leveraging a dynamic graph neural network. However, most of those works mainly focus on user-item interacted graphs. For SBR, TMI-GNN~\cite{Shen2021TemporalAM} proposed to use temporal information to guide the multi-interest network to focus on multi-interest mining. 
TASRec~\cite{DBLP:conf/sigir/ZhouTHZW21} incorporate temporal information by constructing a dynamic graph snapshots sequence at different timestamps. GNG-ODE~\cite{Guo2022EvolutionaryPL} propose graph-nested GRU ordinary differential equation to encode both temporal and structural patterns into continuous-time dynamic embeddings. Orthogonal to these dynamic-based GNNs models, we incorporate material information via a contrastive learning strategy.}

\subsection{Contrastive Learning in RS}

Recently, In the CV and NLP~\cite{wan2023text, xiong2022self, wang2024iot, wan2023efficient} area, multiple contrastive Learning \cite{DBLP:conf/icml/ChenK0H20,  DBLP:journals/corr/abs-2104-02057, gao2021simcse, wang2020understanding, wan2023med, liu2023etp} methods have demonstrated superior performance in modelling representation by measuring the similarity between different views within unlabeled raw data. This self-supervised mechanism is widely adopted in recommendation systems because it carries good semantic or structural meanings and benefits downstream tasks. For instance, GCC \cite{DBLP:conf/kdd/QiuCDZYDWT20} proposed sub-graph instance discrimination that utilized contrastive learning to learn the intrinsic and transferable structural representations.
Yao et al. \cite{DBLP:conf/cikm/YaoYCYCMHCTKE21} proposed multi-task contrastive learning for a two-tower model. Besides, S$^{3}$-Rec \cite{DBLP:conf/cikm/ZhouWZZWZWW20} used mutual information maximization to explore the correlation among items, attributes, and contexts.
Recently, Wei et al. proposed CLCRec \cite{DBLP:conf/mm/WeiWLNLLC21} to leverage contrastive learning to learn the mutual dependencies between item content and collaborative signals in order to solve the cold start problem. Wu et al. \cite{DBLP:conf/sigir/WuWF0CLX21} generated multiple views of the same node from a graph and employed contrastive learning to maximize their agreement to mine hard negative samples.   \bruce{DuoRec~\cite{Qiu2021ContrastiveLF} designed a contrastive regularization to reshape the distribution of sequence representations.}
In the SBR task, Li et al. \cite{DBLP:conf/bigdataconf/LiLYW21} made use of a global-level contrastive learning model to solve noise and sampling problems in heterogeneous graphs. \wan{S$^{2}$-DHCN \cite{DBLP:conf/aaai/0013YYWC021} is the most relevant work to us, which designs a contrastive learning mechanism to enhance hyper-graph modelling via another line GCN model. 
\bruce{COTREC~\cite{Xia2021SelfSupervisedGC} augments the session-based graph with two views that exhibit sessions' internal and external connectivity by contrastive learning.}
\bruce{CORE~\cite{Hou2022CORESA} unifies the representation space of session embeddings using the contrastive-based representation-consistent encoding strategy.}
\bruce{CGL~\cite{Pan2022CollaborativeGL} used the self-supervised learning and main supervised learning to explore the correlations of different sessions for enhancing the item representations.}
But these methods ignore the temporal pattern in the spatial structure, leading to information loss. Orthogonal to these methods, our RESTC employs spatio-temporal contrastive learning to supply sufficient \xin{interactions} between spatial \xin{structures} and temporal \xin{patterns} via aligning the two views in the latent space.}

\section{Overall Framework}
\subsection{Problem Definition}
\label{sec:problem_definition}
\wan{Suppose that the item set is $V = \left\{ v_{1}, v_{2}, \ldots, v_{N}\right\}$, where $v_{i}$ indicates \xin{the $i$-th} item and $N$ is the number of item categories. 
Given an ongoing session denoted as items  \bruce{$s=\left[v_{s_1}, v_{s_2}, v_{s_3}, \ldots, v_{s_M}\right]$}, $v_{s_i} \in V (1 \leq i \leq N)$ represents the $i$-th historical interactive item of the user within session $s$,  and $M$ is the length of the session, it aims to predict the items $v_{M+1}$ that the user will interact with at the next time stamp. Generally, the goal of the session-based recommendation is to recommend the top-K rank items $(1 \leq K \leq N)$ that have the highest probability of being clicked/purchased by the user. }

\subsection{Overview of RESTC} 
\begin{figure*}[htbp]
  \centering
  \includegraphics[width=1.0\textwidth]{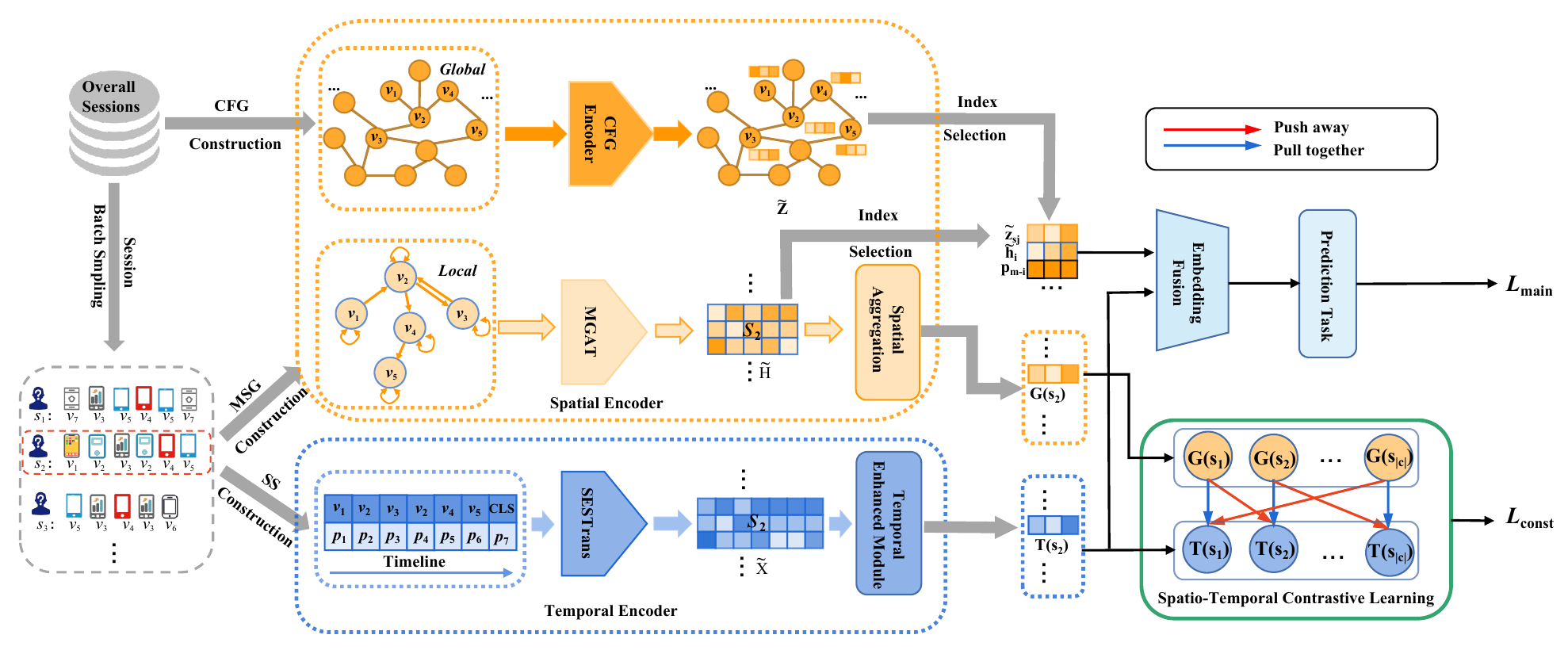}
  \vspace{-0.1in}
  \caption{Overview of RESTC.  MGAT, SESTrans, CFG encoder represent \textit{Multi-relational Graph Attention Network} in Sec~\ref{sec:cgat}, \textit{Session Transformer} in Sec~\ref{sec:SESTrans}, \textit{Collaborative Filtering Graph} encoder in Sec~\ref{sec:cfg_embedding}, respectively.
  }
  \vspace{-0.1in}
  \label{fig:architecture}
\end{figure*}

\bruce{Our RESTC framework includes a \textbf{Spatio-Temporal Contrastive Learning} strategy in Sec~\ref{sec: spatio-temporal contrastive learning} and a \textbf{ Collaborative Filtering Graph} enhanced main supervised task in Sec~\ref{sec:prediction}. 
The training process is shown in Fig~\ref{fig:architecture}: 
First, the session data (e.g., $S_{2}$) is transformed into the two aggregated embeddings ($\mathbf{T}(s)$ and $\mathbf{G}(s)$) encoded by the local spatial and temporal encoders. Then, to remedy the temporal information loss during encoding the spatial structure as shown in Fig~\ref{fig:motivation}, the spatial-temporal contrastive learning is applied to align and interact with the embeddings of the two views in the latent space. Furthermore, In the main prediction task, we enhance the spatial embeddings $\tilde{\mathbf{H}}$ with the Collaborative Filtering Graph (CFG) embedding $\mathbf{\tilde{Z}}$ and apply the embedding fusion to generate session representation to predict the next item. }

\section{Spatio-Temporal Contrastive Learning}
\label{sec: spatio-temporal contrastive learning}
\wan{In this section, we augment a session into two views of embeddings from a \textbf{temporal encoder} in Sec.~\ref{sec:cgat} and a \textbf{spatial encoder}~\ref{sec:SESTrans} respectively.}
To align and interact with the output embeddings from the two encoders in the latent space, we design a contrastive learning task and introduce it in Sec.~\ref{sec:contrastive}.

\subsection{Temporal Encoder for Session Sequences}
\label{sec:SESTrans}
We present how to model session data as sequences from a temporal view, corresponding to the temporal part of Fig.~\ref{fig:architecture}.

\subsubsection{\textbf{Session Sequence Construction}} 

\noindent Given a session \bruce{$s=\left[v_{s_1}, v_{s_2}, v_{s_3}, \ldots, v_{s_M}\right]$}, by adopting an embedding layer, all items in the session will be embedded to a sequence of item embeddings, denoted as $\mathbf{X} = \left[\mathbf{x_{1}},\mathbf{x_{2}}, \mathbf{x_{3}}, \ldots, \mathbf{x_{L}}\right]$. $\mathbf{X} \in \mathbb{R}^{L \times D}$ is the model input. $L$ denotes the max length of all sessions. The zero vector will be padded after the sequence when the length of a session $M$ is shorter than $L$. To help aggregate item embeddings to a fused session representation as a temporal pattern, \bruce{we additionally add the special item [CLS] at the end of the session sequence to learn the global attention information}, similar to BERT-encoder\cite{DBLP:conf/naacl/DevlinCLT19, wan2022g}. To encode temporal information, we equip the initial item embeddings with the learnable absolute temporal position embeddings (denoted as $\mathbf{P_{t}} \in \mathbb{R}^{L\times D}$):

\begin{align}
    \mathbf{X}^{\prime}= \text{Concat} (\bruce{\mathbf{X}}, \mathbf{P_{t}}), \label{eq:E_s}
\end{align}

\noindent where $\mathbf{X}^{\prime} \in \mathbb{R}^{(L+1) \times 2D}$.

\subsubsection{\textbf{Session Transformer Layers for SEs}} 

To obtain preliminary temporal embedding of sessions , we leverage the Session Transformer (SESTrans) following the standard transformer encoder~\cite{NIPS2017_3f5ee243}, which employs weight matrix $\mathbf{W}_{Q}$, $\mathbf{W}_{K}$, $\mathbf{W}_{V}$ to linearly transform the input $\mathbf{X}^{\prime} \in \mathbb{R}^{(L+1) \times 2D} $ as query, key, value vectors, denoted as $\mathbf{Q}$, $\mathbf{K}$, $\mathbf{V}$. The scaled dot-product attention is defined as:

\begin{align}
    \operatorname{Attention}(\mathbf{Q}, \mathbf{K}, \mathbf{V})=\operatorname{softmax}\left(\frac{\mathbf{Q K}^{T}}{\sqrt{2D}}\right) \mathbf{V}. \label{eq:attention}
\end{align}

\nobreak
Intuitively, the attention module aggregates low-level item representations to high-level item representations via a linear combination. We also implement SESTrans in a multi-head fashion like in \cite{NIPS2017_3f5ee243}. Since SAN is linear to input, we feed the output of SESTrans to a feed-forward network (FFN) with non-linearity activation:

\begin{align}
    \operatorname{FFN}(\mathbf{x})=\text{ReLU}\left(\mathbf{x} \mathbf{W_{1}}+\mathbf{b_{1}}\right) \mathbf{W_{2}}+\mathbf{b_{2}}, \label{eq:FFN}
\end{align}

\noindent where $\mathbf{W_{1}}$ and $\mathbf{W_{2}}\in \mathbb{R}^{2D \times 2D}$ and $\mathbf{b_{1}}$, $\mathbf{b_{2}} \in \mathbb{R}^{2D}$ are trainable parameters in FFN layers.
Besides, we stack several encoder layers to learn more complicated session representation from the temporal review, accompanied by standard residual connection, dropout mechanism, and layer normalization. After that, we obtain the encoder's output embedding $\mathbf{\tilde{X}}$.

\subsubsection{\textbf{Temporal Enhanced Module}} 
\label{sec:Generating Temporal View Embedding}

\noindent To better aggregate item embeddings from encoder layers to obtain the user's evolving preference with respect to the timeline, we develop a novel temporal enhanced module. In particular, we utilized the embedding of the special item [CLS] of output embeddings $\mathbf{\tilde{X}}$ as query vector $\mathbf{Q^{\prime}}$, and the rest of output embeddings $\mathbf{\tilde{X}}$ as key vector $\mathbf{K^{\prime}}$. Note that $\mathbf{Q^{\prime}}$ is the global preference representation, and $\mathbf{K^{\prime}}$ is the preference evolution representation. Besides, we leverage initial embedding $\mathbf{X}^{\prime}$ as our value vector $\mathbf{V^{\prime}}$ \wan{since it contains the original temporal positional encoding} information, which can benefit our output embedding with the temporal pattern. Then, we add the two representations and apply a non-linear transformation with ReLU activation. Finally, a softmax function is used to calculate attentive relations and gain the aggregative vector $\mathbf{h_{t}}$.
The formulas are defined as:

\begin{align}
    \operatorname{\mathbf{\gamma_{t}}} &=\text{softmax}\left(\text{ReLU}\left(\mathbf{Q^{\prime} \mathbf{W_{3}}} +\mathbf{K^{\prime} \mathbf{W_{4}}} +\mathbf{b_{3}}\right)\mathbf{f_{t}}\right) , \label{eq:gamma_t} \\
    \operatorname{\mathbf{h_{t}}} &= \sum_{i=1}^{L} \operatorname{{\gamma_{t}}_i}\mathbf{v^{\prime}_{i}}, \label{eq:h_t}
\end{align}

\nobreak
where $\mathbf{W_{3}}, \mathbf{W_{4}}\in \mathbb{R}^{2D\times2D}$ and $\mathbf{f_{t}},\mathbf{b_{3}} \in \mathbb{R}^{2D}$ $\operatorname{\mathbf{\gamma_{t}}}$ is the combined vector.
To this end, we have obtained the aggregation vector $\mathbf{h_{t}}$ and the global preference vector from the embedding of special token [CLS], denoting as $\mathbf{x_{c}}$.
Then we concatenate the two vectors and pass them to a feed-forward layer. Finally, dropout and L2 normalization tricks are employed after the FFN layer then we obtain temporal view embedding as:

\begin{align}
    \mathbf{T}(s) = \operatorname{L2Norm}(\operatorname{FFN}(\text{Concat}(\mathbf{h_{t}}, \mathbf{x_{c}}))). \label{eq:temporal}
\end{align}

\nobreak

\subsection{Spatial Encoder for Session Graphs}
\label{sec:cgat}
The subsection shows the session graph construction process and its learning process, illustrated in the local spatial part of Fig.~\ref{fig:architecture}.

\subsubsection{\textbf{Multi-relational Session Graph Construction}} 
\label{sec:session_graph}

\noindent There may exist duplicate items in one session. Thus, it is important to construct a session graph to capture such the spatial relationship in terms of item transitions.
Given a session $s$ with a \textit{repeatable} item sequence \bruce{$s=\left[v_{s_1}, v_{s_2}, v_{s_3}, \ldots, v_{s_M}\right]$}, 
let $ G_{s} = (V_{s}, E_{s})$ be the corresponding session graph where the node set $V_{s}$ consists the unique items in the session, the edge set $E_{s}$ contains edges represented any two adjacent items $(v_{i}, v_{j})$ in the sequence $s$, forming an item-transition pattern behind the session.
In contrast to FGNN \cite{DBLP:conf/cikm/QiuLHY19} which utilizes the occurrence frequency of edges to construct a weighted directed graph for a session, we leverage a multi-relational weighted graph which uses multiple types of relationship, including in-relation, out-relation, bi-direction and self-loop. Specifically, the out-relation indicates that there only exists a transition $(v_{i}, v_{j})$ in the graph; the in-relation is vice versa. 
The bi-direction represents that $ (v_{i}, v_{j})$ simultaneously exits bi-directional transition. Besides, the self-loop implies that there exist a self transition of an item. By using these four relationships, the spatial structure can be enriched by more accurate inter-relationships among item transitions. We name this graph as Multi-relational Session Graph (MSG).
A concrete example is demonstrated in Fig.~\ref{fig:architecture},
in which the session  $s_1 = [v_{1}, v_{2}, v_{3}, v_{2}, v_{4}, v_{5}]$ can be converted into a multi-relational graph as shown inside the blue dotted rectangle with local.


\subsubsection{\textbf{Multi-relational Graph Attention Network for MSGs}} 
\label{sec:gan}
\noindent We next present how to propagate item features on a multi-relational session graph to encode item-transitional relations. \wan{Graph attention network (GAT)~\cite{DBLP:conf/iclr/VelickovicCCRLB18} and Multi-relational GCN~\cite{DBLP:conf/ijcai/HuangLYN20} have shown their powerful capability in graph structure and multiple types of edge relations learning, respectively.} We further extend them to our multi-relational weighted graph and denote the model as MGAT.
The input to our encoder layer is a set of item features after embedding layer, $\mathbf{H} = \left[\mathbf{h_{1}}, \mathbf{h_{2}}, \mathbf{h_{3}}, \ldots, \mathbf{h_{U}}\right]$, where $\mathbf{h_{i}} \in \mathbb{R}^{D}$, $U$ is the number of  unrepeatable items in current session ($U \le M$), and $D$ is hidden size. We define relation embedding of in-relation, out-relation, bi-direction, and self-loop as  $\mathbf{r_{in}}$, $\mathbf{r_{out}}$, $\mathbf{r_{bi}}$, and $\mathbf{r_{self}}$ respectively.
We denote $\mathbf{r_{ij}}$ as a general relation embedding between $v_i$ and $v_j$ that is determined by the specific relation between the two items, \textit{i.e.},  one of  the four relations.
The attention scores among these items are calculated by 

\begin{align}
    e_{i j} &= \mathbf{r_{ij}}^\intercal \left(\mathbf{h_{i}} \circ \mathbf{h_{j}}\right), \label{eq:e_ij}  \\
    \alpha_{i j} &= \frac{\exp \left(\text{LeakyReLU}(e_{i j})\right)}{\sum_{k \in \mathcal{N}_{i}} \exp  \left(\text{LeakyReLU}(e_{i k})\right)}, \label{eq:alpha_ij}
\end{align}

\nobreak
where $e_{ij}$ is the relational similarity between item $v_{i}$ and its neighbor $v_{j}$ by element-wise product and relational inner product, $\alpha_{i, j}$ is the attention scores.

\wan{It is worth noting that our MGAT is different from~\cite{DBLP:journals/kbs/LiZZZ22,DBLP:conf/iclr/VelickovicCCRLB18, DBLP:conf/sigir/Wang0CLMQ20}, we employ a multi-head attention mechanism to incorporate all edge relations instead of a single head latent space to better enhance the representation ability for the spatial structure}. To be specific, each head computes a kind of relations among items and their neighbors, and then the embeddings of multi-head attention are added rather than concatenated:

\begin{align}
    \mathbf{\tilde{h}_{i}}=\sum_{r=1}^{R} \sum_{j \in \mathcal{N}_{i}} \alpha_{i j}^{(r)} \mathbf{h_{j}}, \label{eq:H_prime}
\end{align}

\nobreak
where $R = 4$ denotes that four relations mentioned above, $\alpha_{i, j}^{(r)}$ are normalized attention coefficients of item $v_{i} $ and its neighbor $v_{j}$ in the $r$-th relation head.
Then, we get the attention-aware representation $\mathbf{\tilde{H}} = \left[\mathbf{\tilde{h}}_{1}, \mathbf{\tilde{h}}_{2},  \mathbf{\tilde{h}}_{3},  \ldots, \mathbf{\tilde{h}}_{M}\right]$ of a specific session based on the initial item order of the session, where $M$ is the item number of the current session.

\subsubsection{\textbf{Local Spatial Aggregation}} 
\label{sec:Generating Spatial View Embedding}
\noindent To emphasize the recent preference within the current session, we concatenate $\mathbf{\tilde{H}}$ representations with a learnable position embedding $\mathbf{P}_{s} = 
\left[\mathbf{p_{M}},\mathbf{p_{m-1}}, \mathbf{p_{m-2}}, \ldots, \mathbf{p_{1}}\right]$.
Besides, the session information can also be represented as the average in general.
Thus, we take the two ways into consideration:

\begin{align}
    \mathbf{\check{H}} &= \operatorname{tanh}(\text{Concat}(\mathbf{P_{s}}, \mathbf{\tilde{H}})\mathbf{W_{s}}), \label{eq:H_star} \\
    \mathbf{\overline{H}_{s}} &= \frac{1}{M} \sum_{i=1}^{M}{\mathbf{\tilde{H}_{i}}}, \label{eq:H_s} \\
    \mathbf{\beta_{s}} &= \operatorname{sigmoid}\left( \mathbf{\check{H}} \mathbf{W_{5}} + \mathbf{\overline{H}_{s}} \mathbf{W_{6}} + \mathbf{b_{5}}\right)\mathbf{f_s}, \label{eq:beta_s}
\end{align}

\nobreak

\noindent where $\mathbf{\check{H}}$ is the position-sensitive session embedding, $\mathbf{\overline{H}_{s}}$ is the average embedding of the general session, $\mathbf{\beta_{s}}$ is soft-attention score indicating the importance of each item, and $\mathbf{W_{s}}\in \mathbb{R}^{2D\times D}, \mathbf{W_{5}}, \mathbf{W_{6}} \in \mathbb{R}^{D \times D}, \mathbf{b_{5}}, \mathbf{f_s} \in \mathbb{R}^{D}$ are trainable parameters.
Finally, the spatial view embedding of a session $s$ is calculated by combing item embeddings with their corresponding importance $\mathbf{\beta_{s}}$:

\begin{align}
    \mathbf{G}(s) = \sum_{i=1}^{M} {\beta_{s}}_i \mathbf{\tilde{h}_{i}}.  \label{eq:spatial}
\end{align}

\subsection{Contrastive Loss function}
\label{sec:contrastive}

One of the key properties of contrastive learning is to align features from positive pairs~\cite{wang2020understanding}. Such positive pairs could be (i) a data sample with two augmentation tricks before being fed into a encoder~\cite{DBLP:conf/icml/ChenK0H20,DBLP:journals/corr/abs-2104-02057}, (ii) a data sample with twice dropout noises in a encoder \cite{gao2021simcse}, or (iii) a data sample with two different encoders~\cite{DBLP:conf/icml/HassaniA20}.
Inspired by the~\cite{DBLP:conf/icml/HassaniA20} which constructs the contrastive samples with two different encoders, \wan{we utilize contrastive learning \xin{to} align the augmented representations from the spatial and temporal encoders in the latent space \xin{and} maximize the lower bound of mutual information of the two views.}

\wan{To achieve the target, we design a spatio-temporal contrastive loss function to distinguish whether the two representations are derived from the same session. Specifically, the contrastive loss learns to minimize the difference between the augmented spatial and temporal views of the same session and maximize the difference between the two augmented views derived from the different sessions. Technically, considering a mini-batch of $C$ sessions $s_{1}$, $s_{2}$, $\ldots$, $s_{i}$, $\ldots$, $s_{C}$, we get the output embeddings from the spatial encoder (see Eq.~\ref{eq:spatial}) and the temporal encoder (see Eq.~\ref{eq:temporal}), denoted as $\mathbf{G}(s_{i})$ and $\mathbf{T}(s_{i})$ for each session, respectively, where we treat $(\mathbf{G}(s_{i}), \mathbf{T}(s_{i}))$ as the positive pair.  For the negative samples, we propose a mixed noise negative sampling strategy that applies a column-wise shuffling operator for each $\mathbf{T}(s_{i})$ in the batch to produce the noisy temporal samples and combine them with all $\mathbf{T}(s)$ to obtain a 2$C$ negative candidate pool, then randomly samples $C$ negative examples denoted as $\mathcal{C}^-$ within the pool. Formally, inspired by SimCLR~\cite{DBLP:conf/icml/ChenK0H20}}, we adopt InfoNCE~\cite{DBLP:journals/jmlr/GutmannH10} as contrastive loss that can be formulated as


\scalebox{0.90}{\parbox{1.05\linewidth}{
\begin{align}
    \mathcal{L}_{cont} = -\sum_{i=1}^{C} \log \frac{\exp \left(\operatorname{sim}\left(\mathbf{G}(s_{i}), \mathbf{T}(s_{i})\right) / \tau \right)}{\sum_{s^{-} \in \mathcal{C}^-} \exp \left(\operatorname{sim}\left(\mathbf{G}(s_{i}), \mathbf{T}(s^{-})\right) / \tau \right)}, \label{eq:L_con}
\end{align}
}}

\nobreak
\noindent \wan{where $\operatorname{sim}(\mathbf{x}, \mathbf{y})=\frac{\mathbf{x}^{\top} \mathbf{y}}{\|\mathbf{x}\|\|\mathbf{y}\|}$ computes the cosine similarity, and $\tau$ is a fixed temperature parameter. By minimize the contrastive objective, we can obtain the enhanced session representations with sufficient \xin{interactions} between spatial and temporal augmented views in the latent space.}

\section{ Main supervised Task of RESTC}
\label{sec:prediction}

Note that the auxiliary contrastive learning task does not need labels.
This section introduces the main supervised task to aggregate spatial and temporal embeddings.
Since collaborative filtering information could also be in the format of graph, we construct the global collaborative filtering graph to enhance the spatial encoder (see details in Sec.~\ref{sec:cfg_embedding}). Sec.~\ref{sec:session_representation} illustrates how to generate the final session representation to fuse the temporal embeddings and the enhanced spatial embeddings, based on which RECTC predicts the next item (see Sec.~\ref{sec:final_prediction}). 
Lastly, Sec.~\ref{sec:multitask} presents how to jointly train the contrastive and downstream tasks via a multi-task fashion.

\subsection{\xin{Spatial Encoder for CFG}}
\label{sec:cfg_embedding}

A  Collaborative Filtering Graph (CFG)
\label{sec:cf_graph} is to learn the collaborative filtering information of a session  based on a global item-transition view. Given a complete session set from all anonymous users, denoted as $S=\left[s_{1}, s_{2}, s_{3}, \ldots, s_{l}\right]$, let $ G_{cf} = (V_{cf}, E_{cf})$ be a graph where $ V_{cf} \in I$ denotes the item set and $E_{cf}$ represents weighted edges from all item-relationships. We define that an item pair has a \textit{connection} in a session if they are adjacent in such a session, the times of repeated \textit{connections} are treated as the weight of the edge between the pair.
\xin{This can be found in the CFG encoder part of Fig.~\ref{fig:architecture}.}


\subsubsection{\textbf{\xin{Collaborative Filtering Graph Encoding}}} 
Obtaining the embedding of CFG enriches a session's representation with implicit collaborative filtering information from other session data. Without the assistance of CFG embeddings, modeling of a single short-term session could be ineffective in capturing complex transitional relationships among items overall sessions, and it will suffer from severe data sparsity problems. In such a case, we leverage the GraphSAGE-GCN~\cite{DBLP:conf/nips/HamiltonYL17}, which used the mean-pooling propagation rule to subtly encode the CFG to aggregate K-hop neighbors' information of every item. The one layer of the encoder is:

\begin{align}
    \mathbf{Z}^{(k)} =\text{LeakyReLU}(\mathbf{\tilde{D}^{-1}} \mathbf{\tilde{A}}  \mathbf{Z}^{(k-1)} \mathbf{W_{c}}^{(k)}), \label{eq:Z}
\end{align}

\nobreak
\noindent where $\mathbf{Z}^{(0)} \in \mathbb{R}^{N\times D}$ represents initial input embedding of items of all sessions, $\mathbf{W_{c}}^{(k)} \in \mathbb{R}^{D \times D}$ denotes learnable weight matrix in the $k$-th layer, $\mathbf{\tilde{A}}=\mathbf{A}+\mathbf{I_{N}}$ means that adjacent matrix added with identity matrix, which can be seem as a self-loop of items in CFG. And $\mathbf{\tilde{D}_{i i}}=\sum_{j} \mathbf{\tilde{A}_{i j}}$ are degree matrix over CFG. After passing $K$ layers graph convolution encoder, we get the K-hop CFG embedding represented as $\mathbf{\tilde{Z}} = \mathbf{Z}^{(K)} = \left[\mathbf{z_{1}}^{(K)}, \mathbf{z_{2}}^{(K)}, \mathbf{z_{3}}^{(K)}, \ldots, \mathbf{z_{N}}^{(K)}\right]$, where $N$ is the number of items overall sessions.

\subsubsection{\textbf{\xin{Spatial Encoder Enhancing with CFG embedding}}} 

We additionally add the K-hop neighbor view from CFG  (denoted as $\mathbf{\tilde{Z}}$) to obtain the enhanced graph-structure representation, which is extracted from global CFG embeddings that involve items in the current session $s$ (denoted as $\mathbf{\tilde{Z}_{s}}$). The embedding of a specific session is: 

\begin{align}
    \mathbf{H_{g}}= \text{Concat}(\mathbf{P_{e}}, \mathbf{\tilde{H}}, \mathbf{\tilde{Z}_{s}}) \mathbf{W_{g}},  \label{eq:Hg_concat}
\end{align}

\nobreak

where $\mathbf{W_{g}}\in \mathbb{R}^{3D\times D}$ is trainable parameter, $\mathbf{P_{e}}$ is the position embedding mentioned in Eq.\ref{eq:H_star}, $\mathbf{\tilde{H}}$ is the output embedding of MSG in Eq.\ref{eq:H_prime}. To this end, we have obtained enhanced graph-based session embedding that simultaneously contains the spatial view of the current session and global collaborative filtering from all sessions.

\subsection{\xin{Embedding Fusion of The Two Views}}
\label{sec:session_representation}
After the session data pass through the encoders from the spatial and the temporal views at the meantime, we obtain the distinct semantic representations from the two views. 
To generate the hybrid preference representation considering both the advantages of each view, we also apply the soft-attention mechanism to combine the enhanced spatial graph embeddings with temporal embeddings to acquire attentive vectors $\rho_{s}$ of each item.
The details are listed as follows:

\begin{align}
    \mathbf{H_{g}^{\prime}} &= \operatorname{tanh}(\mathbf{H_{g}}\mathbf{W_{f}}), \label{eq:Hg_prime_tanh} \\
    \mathbf{\rho_{s}} &= \operatorname{sigmoid}\left( \mathbf{H_{g}^{\prime} \mathbf{W_{7}}}+ \mathbf{T} \mathbf{W_{8}}+ \mathbf{b_{7}}\right)\mathbf{f_g}, \label{eq:alpha_sigmoid} \\
    \mathbf{s_{h}} &= \sum_{i=1}^{U}{{\rho_{s}}_{i}\left(\mathbf{{\tilde{z}}_{s_i}} + \mathbf{\tilde{h}_{i}}\right)}, \label{eq:Sh_sum}
\end{align}

\nobreak

where $\mathbf{H}_{g}$ is the spatial embedding from Eq. (\ref{eq:Hg_concat}), $\mathbf{T}$ is the temporal embedding from Eq. \ref{eq:temporal}, $\mathbf{{\tilde{z}_{s_i}}}$ indicates the CFG embedding of the $v_i$ in session $s$, and $\mathbf{\tilde{h}_{i}}$ denotes the MSG embedding of $v_i$, $\mathbf{W_{f}}, \mathbf{W_{7}}, \mathbf{W_{8}} \in \mathbb{R}^{D\times D}$ are learnable matrices, and $\mathbf{b_{7}}, \mathbf{f_g} \in \mathbb{R}^{D}$ are learnable biases.
Finally, we get the semantic-rich representation $\mathbf{s_{h}}$ which \xin{incorporates} \wan{the global collaborative filtering spatial, the session spatial, and the session temporal information.}

\subsection{Next-item Prediction Task}
\label{sec:final_prediction}

We further make use of the session embedding $\mathbf{S_{h}}$ to make recommendations by computing the probability distributions of the candidate items. Specifically, we utilize the softmax function to obtain the main task output:

\begin{align}
    \hat{\mathbf{y}}=\operatorname{softmax}\left(\mathbf{s_{h}} \mathbf{W}_{y} \right), \label{eq:y_hat}
\end{align}

\nobreak
\noindent where $\mathbf{W}_{y} \in \mathbb{R}^{D \times N}$ is transformation matrix for the distribution prediction, $\hat{\mathbf{y}}_{i}$ represent the output probability of the prediction. Then, we apply cross-entropy as our objective function of the main task with the ground truth $\{\mathbf{y}_{1}, \mathbf{y}_{2}, \mathbf{y}_{3}, \ldots, \mathbf{y}_{N}\}$:

\begin{align}
    \mathcal{L}_{main}=-\sum_{i=1}^{N} \mathrm{y}_{i} \log \left(\hat{\mathbf{y}}_{i}\right)+\left(1-\mathrm{y}_{i}\right) \log \left(1-\hat{\mathbf{y}}_{i}\right) \label{eq:L_main}
\end{align}

\begin{table}
\caption{Dataset Statistics.}
\vspace{-0.1in}
\label{tab:data}
\scalebox{0.75}{
  \begin{tabular}{cccccc}
    \midrule[1.2pt]
    \bfseries Dataset & \bfseries Items & \bfseries Clicks & \bfseries Train & \bfseries Test & \bfseries Avg.len\\
    \midrule
    Tmall        & $\hat{\mathbf{y}}=\operatorname{softmax}\left(\mathbf{s_{h}} \mathbf{W}_{y} \right), \label{eq:y_hat}$ & 818,479 & 351,268 & 25,898 & 6.69\\
    Diginetica   & 43,097 & 982,961 & 719,470 & 60,858 & 5.12\\
    Gowalla      &  29,510 & 1,122,788 & 419,200 & 155,332 & 3.85\\
    RetailRocket & 36,968 & 710,586 & 433,648 & 15,132 & 5.43\\
    Nowplaying   & 60,417 & 1,367,963 & 825,304 & 89,824 & 7.42\\
    LastFM       & 38,615 & 3,835,706  & 2,837,330 & 672,833 & 11.78\\
  \midrule[1.2pt]
\end{tabular}
}
\vspace{-0.1in}
\end{table}

\subsection{Multi-task Training for Contrastive and Supervised Tasks}
\label{sec:multitask}

We unify the main recommendation task with the contrastive learning task to enhance the performance of SBR, which could be viewed as a multi-task training process:
\begin{equation}
    \mathcal{L}=\mathcal{L}_{\text {main }}+\eta_{1} \mathcal{L}_{cont}+\eta_{2}\|\Theta\|_{2}^{2},
\end{equation}
where $\eta_{1}$ controls the strength of contrastive learning and $\eta_{2}$ is the constant of $L_{2}$ regularization of the all trainable parameters $\Theta$. Finally, the whole training procedure of RESTC is summarized in Algorithm 1.

\begin{algorithm}[htbp]  \label{ag:core}
\small
  \caption{Training Process of RESTC}
  \begin{algorithmic}[1]\label{ag:core}
    \Require
      Sessions $\mathbf{S}$, item embeddings $\mathbf{V_{s}}$ 
    \Ensure
      Top-k recommendation items
    \State  Transform session data into spatial and temporal view 
     \State Construct CFG overall sessions

    \For{epoch in range(Epoches)}
      \For{batch in DataLoader}
         \For{each session s in batch}
         \State \textbf{Spatio-Temporal Contrastive Learning task:}
        \State Spatial view embedding $\mathbf{G}(s) \gets$ Eq.(1) to (6)
        \State Temporal view embedding $\mathbf{T}(s) \gets$ Eq.(7) to (13)
        \State Contrastive loss $\mathcal{L}_{cont} \gets$ Eq.(14)
        \State \textbf{Prediction task:}
        \State CFG embedding $\mathbf{\tilde{Z}} \gets$ Eq.(15)
        \State Enhanced spatial embedding $\mathbf{H_{g}} \gets$ Eq.(16) 
        \State Embedding Fusion $\mathbf{S_{h}} \gets$ Eq.(17) to (19)
        \State Next-item Predition loss $\mathcal{L}_{main} \gets$ Eq.(20), (21)
        
          \EndFor
        \EndFor
        \State $\mathcal{L}=\mathcal{L}_{\text {main}}+\eta_{1} \mathcal{L}_{cont}+\eta_{2}\|\Theta\|_{2}^{2}$
        \State Using multi-task training to jointly optimize $\mathcal{L}$
    \EndFor
  \end{algorithmic}
\end{algorithm}

\section{Experiment}

\subsection{Experimental Settings}
In this section, aiming to answer the following research question, we conduct extensive experiments on six datasets.

\begin{itemize}[leftmargin=*]

    \item \textbf{RQ1} How does RESTC perform compared to present methods in the SBR task?
    \item \textbf{RQ2} Are the main components (e.g., Session graph encoder (MGAT), Temporal encoder (SESTrans), CFG encoder, spatio-temporal contrastive learning) really working well?
    \item \textbf{RQ3} How does the spatial encoder (MGAT) work effectively compared to other GNN-based backbones?
    \item \textbf{RQ4} How do different settings (temperature $\tau$, negative sampling strategies) of contrastive learning impact the performance of RESTC?
     \item \textbf{RQ5} Are RESTC robust to different lengths of session data? 
     \item \textbf{RQ6} How do different hyper-parameters affect RESTC?
      \item \textbf{RQ7} Is the spatio-temporal contrastive learning really improving the representation learning?

\end{itemize}

\subsubsection{\textbf{Dataset Description}} 

\noindent We evaluate our RESTC on six public benchmark datasets: {\itshape Tmall\footnote{https://tianchi.aliyun.com/dataset/dataDetail?dataId=42}}, {\itshape Diginetica\footnote{http://2015.recsyschallenge.com/challenge.html}}, {\itshape Gowalla\footnote{https://snap.stanford.edu/data/loc-gowalla.html}}, {\itshape RetailRocket\footnote{https://www.kaggle.com/retailrocket/ecommerce-dataset}}, {\itshape Nowplaying\footnote{http://dbis-nowplaying.uibk.ac.at/}}, {\itshape LastFM\footnote{http://ocelma.net/MusicRecommendationDataset/lastfm-1K.html}}, which are often used in session-based recommendation models.
\textbf{Tmall}  comes from a competition in IJCAI, which contains anonymous users' shopping logs on the Tmall online website.
\textbf{Diginetica}  records the clicks of anonymous users within six months, and it is from the CIKM Cup platform 2016. 
\textbf{Gowalla} is a check-in dataset that is widely utilized by point-of-interest recommendation. We follow \cite{DBLP:conf/kdd/ChenW20} to process this data.
\textbf{RetailRocket}   is original from a Kaggle contest published by an e-commerce company, which contains the browser activity of anonymous users within six months.
\textbf{Nowplaying}   describes the music listening behavior of users, and it comes from the resource of \cite{DBLP:conf/mm/ZangerlePGS14}.
\textbf{LastFM: }  is a popular music dataset that has been used as a benchmark in many recommendation tasks. Following \cite{DBLP:conf/kdd/GuoYWCZH19}, we employ it as session-based data. 

Moreover, we adopt the data augmentation and filtering for the sessions following by~\cite{DBLP:conf/aaai/WuT0WXT19, DBLP:conf/cikm/QiuLHY19,DBLP:conf/ijcai/XuZLSXZFZ19,DBLP:conf/ijcai/LuoZLZWXFS20}. Specifically, we process these datasets into sessions. Concretely, we get rid of all sessions whose length is shorter than 1 and the appearing of items less than 5 overall sessions. We also set the data of last 7 days to be the test data and the previous data as train data. In addition, given a session data $s=\left[v_{1}, v_{2}, \ldots, v_{M}\right]$, we augment the sequence and generate corresponding labels by splitting it into $([v_{1}], v_{2}), ([v_{1}, v_{2}], v_{3})$, $\ldots, ([v_{1}, v_{2}, $ $\ldots, v_{\xin{M}-1}], v_{M})$ for all sessions in six datasets. The details of processed data are shown in Table~\ref{tab:data}.

\subsubsection{\textbf{Baselines}} \bruce{For fair comparisons, we compare RESTC with sequential-based methods, GNN-based methods, temporal-augmented methods and contrastive learning methods (e.g., self-supervised sequential methods~\cite{https://doi.org/10.48550/arxiv.2010.14395, Qiu2021ContrastiveLF, Hou2022CORESA} and graph contrastive learning approaches~\cite{DBLP:conf/aaai/0013YYWC021, Xia2021SelfSupervisedGC}), respectively.}

\noindent \textbf{Sequential-based Methods: }
\begin{itemize}[leftmargin=*]
    \item \textbf{FPMC}~\cite{DBLP:conf/www/RendleFS10} learns the representation of session via Markov-chain based method. We ignore the user profile information in the experiment and adapt it to the session-based recommendation. 
    
    \item \textbf{GRU4Rec}~\cite{DBLP:journals/corr/HidasiKBT15} is an RNN-based method that utilizes GRU and adopts ranking-based loss to the model preference of users within the current session.
    
     \item \textbf{NARM}~\cite{DBLP:conf/cikm/LiRCRLM17} is a attention-based RNN model to learn session embedding. 
     
     \item \textbf{STAMP}~\cite{DBLP:conf/kdd/LiuZMZ18} is an attention model to capture user’s temporal interests from historical clicks in a session and relies on self-attention of the last item to represent users’ short-term interests.

     \item \bruce{\textbf{SASRec}~\cite{Kang2018SelfAttentiveSR} is a self-attention-based sequential recommendation model that allows us to capture long-term semantics.}

     \item \bruce{\textbf{BERT4Rec}~\cite{Sun2019Bert4rec} employs the deep bidirectional self-attention to model user behaviour sequences.}

\end{itemize}

\noindent \textbf{GNN-based Methods: }
\begin{itemize}[leftmargin=*]
     \item \textbf{SR-GNN}~\cite{DBLP:conf/aaai/WuT0WXT19} is the first GNN-based model for the SBR task, which transforms the session data into a direct unweighted graph and utilizes gated GNN to learn the representation of the item-transitions graph.
     
     \item \textbf{GC-SAN}~\cite{DBLP:conf/ijcai/XuZLSXZFZ19} uses gated GNN to extract local context information and then employs the self-attention mechanism to obtain the global representation.
     
     \item \textbf{CSRM}~\cite{DBLP:conf/sigir/WangRMCMR19} integrates an internal memory encoder through an external memory network by considering the correlation between neighboring sessions.
     
     \item \textbf{FGNN}~\cite{DBLP:conf/cikm/QiuLHY19} proposes to leverage a weighted graph attention network for computing the information flow in the session graph and generates the user preference by a graph readout function.
     
     \item \wan{\textbf{GCE-GNN}~\cite{DBLP:conf/sigir/Wang0CLMQ20} transforms the sessions into global graph and local graphs to enable cross session learning.} 
\end{itemize}   

\noindent \textbf{Temporal-Augmented Methods: }

\begin{itemize}[leftmargin=*]
     \item \wan{\textbf{TASRec}~\cite{DBLP:conf/sigir/ZhouTHZW21} incorporates temporal information via  constructing a sequence of dynamic graph snapshots at different timestamps.}

       \item \bruce{\textbf{TMI-GNN}~\cite{Kumar2019PredictingDE} leverages temporal information to guide
the multi-interest network to focus on multi-interest mining.}

       \item \bruce{\textbf{GNG-ODE}~\cite{Guo2022EvolutionaryPL} uses graph-nested GRU ordinary differential equation to encode both temporal and structural patterns into continuous-time dynamic embeddings.}
\end{itemize} 

\noindent \textbf{Contrastive Learning Methods: }

\begin{itemize}[leftmargin=*]
   \item \bruce{\textbf{CL4Rec}~\cite{https://doi.org/10.48550/arxiv.2010.14395} uses contrastive learning to learn the mutual dependencies between item content and collaborative signals.}

   \item \bruce{\textbf{DuoRec}~\cite{Qiu2021ContrastiveLF} designs a contrastive regularization to reshape the distribution of sequence representations.}

   \item \bruce{\textbf{CORE}~\cite{Hou2022CORESA} unifies the representation space of session embeddings by using the contrastive-based representation-consistent encoding strategy.}

     \item \textbf{$\mathbf{S^{2}}$-DHCN} \cite{DBLP:conf/aaai/0013YYWC021} transforms the session data into hyper-graph and line-graph and and uses self-supervised learning to enhance session-based recommendation.

     \item \bruce{\textbf{COTREC}~\cite{Xia2021SelfSupervisedGC} exploits the session-based graph to augment two views that exhibit the internal and external connectivity of sessions by contrastive learning.}
\end{itemize}  

\begin{table*}[h]\small
\setlength\tabcolsep{2pt}
\caption{Comparisons over all datasets.}
\vspace{-0.05in}
\scalebox{0.7}{
\begin{tabular}{c|c|ccccccccccccc|c}
\midrule[1.2pt]
Dataset & Metric  & FPMC & GRU4REC & NARM & STAMP & \bruce{SASRec} & \bruce{BERT4Rec} & SR-GNN & CSRM & FGNN & GC-SAN & GCE-GNN & TASRec & S$^{2}$-DHCN & $\mathbf{RESTC}^{\star}$ \\

\midrule
\multirow{4}*{\textbf{TM}} 

& HR@10 & $13.10$ & $9.47$ & $19.17$ & $22.63$ & 21.91 & 22.38 & $23.41$ & $25.54$ & $20.67$ & $24.78$ & $ \underline{28.01}$ & $25.72 $ &$26.22$ & $\mathbf{35.57}$  \\

& HR@20  & $16.06$ & $10.93$ & $23.30$ & $26.47$ & 27.72 & 28.12 & $27.57$ & $29.46$ & $25.24$ & $28.72$ & $\underline{33.42}$ & $29.58$ &$31.42$ & $\mathbf{42.47}$  \\

& MRR@10 & $7.12$ & $5.78$ & $10.42$ & $13.12$ & 11.25 & 11.58 & $13.45$ & $13.62$ & $10.67$ & $13.55$ & $\underline{15.08}$ & $14.22$ &  $14.60$ & $\mathbf{18.05}$  \\

& MRR@20 & $7.32$ & $5.89$ & $10.70$ & $13.36$ & 12.11 & 12.85 & $13.72$ & $13.96$ & $10.39$ & $13.43$ & $\underline{15.42}$ & $14.51$ &  $15.05$ & $\mathbf{18.52}$  \\
\midrule
\multirow{4}*{\textbf{DG}} 

& HR@10  & $15.43$ & $17.93$ & $35.44$ & $33.98$ & 35.84 & 36.78 & $36.86$ & $36.59$ & $37.72$ & $37.86$ & $\underline{41.16}$ & 39.85 & $40.21$ & $\mathbf{42.35}$ \\

& HR@20 & $26.53$ & $29.45$ & $49.70$ & $45.64$ & 48.78 & 50.12 & $50.73$ & $50.55$ & $50.58$ & $50.84$ & $\underline{54.22}$ &52.53&$53.66$ & $\mathbf{55.93}$  \\

& MRR@10 & $6.20$ & $7.33$ & $15.13$ & $14.26$ & 14.55 & 15.61 & $15.52$ & $15.41$ & $15.95$ & $16.89$ & $\underline{18.15}$ &17.19& $17.59$ & $\mathbf{18.75}$ \\

& MRR@20 & $6.95$ & $8.33$ & $16.17$ & $14.32$ & 17.22 & 17.16 & $17.59$ & $16.38$ & $16.84$ & $17.79$ & $\underline{19.04}$ &18.22& $18.51$ & $\mathbf{19.65}$ \\


\midrule
\multirow{4}*{\textbf{RR}} 

& HR@10 & $25.99$ & $38.35$ & $42.07$ & $42.95$ & 37.55 & 38.92 & $43.21$ & $43.47$ & $43.75$ & $43.53$ & $\underline{48.22}$ & 46.32 & $46.15$ & $\mathbf{50.12}$ \\

& HR@20 & $32.37$ & $44.01$ & $50.22$ & $50.96$ & 45.85 & 46.72 & $50.32$ & $51.02$ & $50.99$ & $50.71$ & $\underline{55.78}$ & 54.23 & $53.66$ & $\mathbf{57.81}$  \\

& MRR@10 & $13.38$ & $23.27$ & $24.88$ & $26.41$ & 22.12 & 23.44 & $26.07$ & $25.58$ & $26.11$ & $26.03$ & $\underline{28.36}$ & 27.22 & $26.85$ & $\mathbf{30.15}$ \\

& MRR@20 & $13.82$ & $23.67$ & $24.29$ & $25.17$ & 23.39 & 25.54 & $26.57$ & $26.19$ & $26.21$ & $25.76$ & $\underline{28.72}$ & 28.37 & $27.30$ & $\mathbf{30.82}$  \\

\midrule
\multirow{4}*{\textbf{LF}} 

& HR@10  & $6.65$ & $11.21$ & $15.37$ & $14.99$ & 14.44 & 14.72 & $15.12$ & $15.47$ & $15.32$ & $15.68$ & $\underline{17.22}$ & 16.83 & $17.09$ & $\mathbf{18.57}$ \\

& HR@20  & $12.91$ & $17.79$ & $21.86$ & $22.06$ & 19.52 & 20.22 & $22.29$ & $22.31$ & $22.18$ & $22.64$ & $\underline{24.05}$ & 23.22 & $22.86$ & $\mathbf{25.54}$  \\

& MRR@10  & $3.21$ & $4.79$ & $7.12$ & $7.27$ & 6.95 & 7.11 & $7.19$ & $7.33$ & $7.09$ & $7.62$ & $ 7.74$ & $\underline{8.22}$ & $8.02$& $\mathbf{8.87}$  \\

& MRR@20 & $3.73$ & $5.41$ & $7.55$ & $7.84$ & 7.22 & 7.38 & $8.31$ & $8.12$ & $8.03$ & $8.42$ & $ 8.19$ & $\underline{8.65}$ & $8.45$ & $\mathbf{9.28}$ \\

\midrule
\multirow{4}*{\textbf{NP}} 

& HR@10  & $5.28$ & $6.74$ & $13.60$ & $13.22$ & 14.19 & 14.78 & $14.17$ & $13.20$ & $13.89$ & $14.11$ & 16.94 & 16.35 &$\underline{17.35}$ & $\mathbf{18.39}$ \\

& HR@20  & $7.36$ & $7.92$ & $18.59$ & $17.66$ & 20.53 & 21.21 & $18.87$ & $18.14$ & $18.75$ & $19.19$ & 22.37 & 20.52 &$\underline{23.50}$ & $\mathbf{24.79}$ \\

& MRR@10  & $2.68$ & $4.40$ & $6.62$ & $6.57$ & 6.88 & 7.01 & $7.15$ & $6.08$ & $6.80$ & $7.11$ & $\underline{8.03}$ &7.37& $7.87$ & $\mathbf{8.31}$ \\

& MRR@20 & $2.82$ & $4.48$ & $6.93$ & $6.88$ & 7.85 & 8.11 & $7.74$ & $6.42$ & $7.15$ & $7.54$ & $\underline{8.40}$ &7.78&$8.18$ & $\mathbf{8.72}$\\

\midrule
\multirow{4}*{\textbf{GW}} 

& HR@10  & $20.47$  & $31.56$ & $40.53$ & $40.99$ & 38.45 & 39.22 & $41.89$ & $42.11$ & $42.09$ & $42.17$ & $44.25$ & 43.21 &$\underline{45.11}$ & $\mathbf{47.86}$  \\

& HR@20  & $29.91$ & $41.91$ & $50.11$ & $50.15$ & 48.53 & 49.12 & $50.29$ & $50.17$ & $50.11$ & $50.71$ & $52.48$ & 53.55 & $\underline{53.34}$ & $\mathbf{56.38}$ \\

& MRR@10  & $9.88$ & $17.85$ & $22.94$ & $23.10$ & 22.18 & 22.95 & $23.78$ & $23.33$ & $22.91$ &$23.77$ & $\underline{24.11}$ & 23.19 & $23.29$ & $\mathbf{25.33}$ \\

& MRR@20  & $11.37$ & $18.29$ & $23.89$ & $24.03$ & 22.52 & 23.35 & $24.31$ & $24.23$ & $24.11$ & $24.58$ & $\underline{24.68}$ & 23.73 & $23.88$ & $\mathbf{25.92}$  \\
\midrule[1.2pt] 
\end{tabular}
}\label{tab:full_result}

\begin{tablenotes}
        \setlength\labelsep{0pt}
	\begin{footnotesize}
	\item
          $\star$ indicates a statistically significant level $p$-value $<$0.001 comparing our RESTC with the baselines. Underlined numbers mean best baseline. The best performance for each benchmark is marked in black bold. TM, DG, RR, LF, NP, GW denote Tmall, Dignetica, RetailRocket, LastFM, Nowplaying and Gowalla, respectively.
        \par
	\end{footnotesize}
\end{tablenotes}

\vspace{-0.2in}

\end{table*}

\subsubsection{\textbf{Evaluation Merics and Parameter Settings}} 

\noindent Following the baselines mentioned above, we adopt two widely used metrics for the SBR task: \textbf{HR@N} (Hit Rate) and \textbf{MRR@N} (Mean Reciprocal Rank). We report their optimal performance for each baseline following the original setting from their papers. In our settings, we apply grid search to find the optimal parameters based on the random 20$\%$ of train data as validation. Concretely, we search the embedding dimension from the range $\left\{100, 150, 200, 250, 300, 350 \right\}$, and the default batch size is set to 512. We also investigate the coefficient of the contrastive learning task from 5e-4 to 1e-1. In our experiments, the default constant of $L_{2}$ regularization is 1e-5. We stack 2 SESTrans encoder layers as default, which achieve the best performance to capture the temporal patterns in our experiments. Then we search the  MGAT  and  CFG encoding layers from 1 to 4, \wan{we find that 1 MGAT layer and 3 CFG embedding layers \xin{are} already enough for learning the spatial structure representation of a session.} Besides, we utilize the Adam optimizer with a learning rate of 0.001 as well as Step-LR and Cosine-Annealing-LR schedulers to adjust the learning rate. 
\xin{More experimental details are shown on Sec~\ref{sec: hyparameter_analysis}. }

\subsection{Overall Results (RQ1)}

The experiment results of baselines and RESTC model over six datasets are reported in Table~\ref{tab:full_result}. The performance results show that the traditional machine learning method FPMC is worse than deep learning methods since it cannot capture long-time dependency. For sequence-based methods, STAMP and NARM perform better than GRU4REC since they utilize attention mechanisms to learn the critical relations among all items. Besides, CSRM performs the best among sequence-based baselines, demonstrating the effectiveness of leveraging collaborative filtering information from other sessions. Besides, CSRM performs the best compared with STAMP and NARM, demonstrating the efficacy of leveraging collaborative filtering information from other sessions. 

Note that GNN-based methods outperform sequence-based methods, which indicates that there still exists some functional yet undiscovered spatial-structure patterns in sequence-based methods; Moreover, information on item-transition graphs (in the spatial view) might be relatively more informative than the temporal view as in sequence-based methods. Specifically, GC-SAN shows better results than SR-GNN, demonstrating that combining GNN with self-attention could better model the current session’s local and global context information. \wan{GCE-GNN shows better results than SR-GNN and GC-SAN, demonstrating that combining the information of \xin{local sessions and the global neighbor graph} effectively enriches the session representation. TASRec outperforms general GNN-based methods like SR-GNN, FGNN, and GC-SAN, proving that incorporating temporal information is significant to spatial structure. S$^{2}$-DHCN shows excellent performance in LastFM and Gowalla in terms of HR@20} since it uses inter- and intra-relations overall sessions and then applies self-discrimination to improve the representation.

\begin{table}[h]\small
\caption{\bruce{Comparisons with contrastive learning and temporal augmented methods.}}
\vspace{-0.05in}
\scalebox{0.8}{
\begin{tabular}{c|c|ccc|ccc|cc|c}
\midrule[1.2pt]
Dataset & Metric & CL4Rec & DuoRec & CORE & TASRec & TMI-GNN  & GNG-ODE & S$^{2}$-DHCN & COTREC & $\mathbf{RESTC}^{\star}$ \\

\midrule
\multirow{2}*{\textbf{TM}} 

& HR@20 &  $31.75$ & $33.21$ & $32.44$ & $29.58$ & $36.55$ & $\underline{37.68}$ & $31.42$ & $35.19$ & $\textbf{42.47}$ \\

& MRR@20  & $13.33$ & $14.38$ & $13.99$ & $14.51$ & $16.78$ & $\underline{17.88}$ & $15.05$ & 17.55 & $\textbf{18.52}$ \\
\midrule
\multirow{2}*{\textbf{NP}} 

& HR@20 &  $20.66$ & $22.44$ & $21.35$& $20.52$ & $21.86$ & $22.43$ & $\underline{23.50}$ & $22.39$ &\textbf{24.79} \\

& MRR@20  & $7.88$ & $8.31$ & $7.66$ &$7.78$ & $8.22$ & $\textbf{8.77}$ & $8.18$ & $8.31$ & $\underline{8.72}$ \\


\midrule
\multirow{2}*{\textbf{DG}} 

& HR@20  & $49.85$ & $50.25$ & $51.12$ & $52.23$ & $51.84$ & $52.09$ & $53.66$ & $\underline{53.85}$ & \textbf{55.93} \\

& MRR@20 & $17.33$ & $17.88$ & $18.19$ & $18.22$ & $17.98$ & $18.55$ & $18.51$ & $\underline{18.82}$ & \textbf{19.65} \\

\midrule[1.2pt] 
\end{tabular}
}\label{tab:additional_comparision}

\begin{tablenotes}
        \setlength\labelsep{0pt}
	\begin{footnotesize}
	\item
           $\star$ indicates a statistically significant level $p$-value $<$0.001 comparing our RESTC with the baselines. The best performance for each benchmark is marked in black bold. Underlined numbers mean the second best result. We concisely select TM, NP and DG as the datasets, HR@20 and MRR@20 as the metric to compare our RESTC with advanced temporal-augmented, supervised sequential and graph contrastive learning approaches.
        \par
	\end{footnotesize}
\end{tablenotes}

\end{table}

\bruce{For our RESTC, the results show that it significantly outperforms most of the comparative baselines, including sequence-based, GNN-based, temporal-augmented, and \bruce{supervised sequential methods and graph contrastive learning methods as shown in Table~\ref{tab:full_result} and Table~\ref{tab:additional_comparision}}. Especially the comparison on six benchmarks, RESTC show better results than other baselines on TM, DG, RR, LF and GW, which reflects RESTC's superior representation capability. In particular, the significant improvement of RESTC over strong supervised sequential baselines (e.g., CL4Rec, DuoRec and CORE) and graph contrastive learning baselines (e.g., S$^{2}$-DHCN and COTREC), implies that leveraging temporal and collaborative filter information is potential for refining the session representation. Besides, RESTC outperforms temporal-augmented GNNs like TASRec, TMI-GNN and GNC-ODE on most settings, indicating that adequate interactions between spatial and temporal views via contrastive learning can significantly boost performance. Therefore, the comparative results demonstrate the effectiveness and generalization ability of our RESTC framework.}

\begin{table}[htbp]
\caption{Ablation Study in Variants of RESTC.}
\vspace{-0.1in}
 \scalebox{0.75}{
  \begin{tabular}{l|cc|cc|cc}
\midrule[1.2pt]
Dataset & \multicolumn{2}{|c|}{ \textbf{TM} } & \multicolumn{2}{c|}{ \textbf{DG} } & \multicolumn{2}{c}{ \textbf{RR} } \\
\midrule
Measures & \multicolumn{2}{|c|}{ HR@20 MRR@20 } & \multicolumn{2}{c|}{ HR@20 MRR@20 } & \multicolumn{2}{c}{ HR@20 MRR @20 } \\
\midrule
 w/o SESTrans  & $36.61$ & $16.08$ & $54.31$ & $19.11$ & $52.65$ & $27.11$ \\
 w/o CFG & $36.96$ & $16.38$ & $54.28$ & $18.84$ & $56.28$ & $28.94$ \\
 w/o Cont. & $39.27$ & $17.11$ & $54.65$ & $19.35$ & $57.01$ & $29.13$ \\
\midrule
w/o PE-G & $40.81$ & $17.75$ & $51.98$ & $18.04$ & $53.51$ & $27.96$ \\
w/o PE-S & $41.05$ & $17.92$ & $54.45$ & $19.29$ & $56.98$ & $30.03$ \\
\midrule
RESTC & $\mathbf{42.47}$ & $\mathbf{18.52}$ & $\mathbf{55.93}$ & $\mathbf{19.65}$ & $\mathbf{57.81}$ & $\mathbf{30.82}$ \\
\midrule[1.2pt]
\end{tabular}}\label{tab:ablation}
\vspace{-0.2in}
\end{table}

\subsection{\bruce{Ablation Study}}
\subsubsection{\bruce{\textbf{Ablation on main Components (RQ2).}}}
 We further investigate the effectiveness of each module in our RESTC model by conducting Ablation experiments. Concretely, we design several contrast variants of RESTC, and they are: (i) w/o SESTrans, which removes the temporal encoder SESTrans thus without the spatio-temporal contrastive learning; (ii) w/o CFG, which only considers the spatial encoder MGAT and spatial encoder SESTrans, without the CFG embedding; (iii) w/o Cont, which contains two complete augmented encoders without the contrastive learning task. Besides, to investigate the impact of position embedding for the spatial and temporal views, (iv) w/o PE-G and w/o PE-S represent RESTC model without learnable position embedding in spatial encoder MGAT and without timeline absolute position embedding in temporal encoder SESTrans.

From Table~\ref{tab:ablation}, we can observe that removing the above components consistently leads to a performance drop, implying that these components are all significant to RESTC. Concretely,  w/o SESTrans underperforms w/o Cont, showing that incorporating temporal information through directly combining the temporal embedding with spatial embedding in the main supervised task has already improved the performance. \wan{Then, the downward trend of w/o CFG is more evident than w/o Cont. The phenomenon is consistent with our assumption that obtaining the implicit collaborative filtering information from the global weighted session graph, denoted as CFG, can enhance spatial representation, which can help remedy the data sparsity problem for the short-term session.} \wan{Furthermore, it can be observed that spatio-temporal contrastive learning enhances the performance on both metrics by comparing standard RESTC with RESTC w/o Cont with an obvious margin. This reveals that cross-view \xin{interactions} via contrastive regularization in the latent space can further reinforce the session representation for the main prediction task.}

Besides, w/o PE-G demonstrates that removing the position embedding in the spatial structure view results in a remarkable performance drop since the model cannot recover the initial order relation after graph embedding.  Moreover, w/o PE-S performs worse than RESTC in the selective datasets and shows the effectiveness of temporal-aware encoding in the temporal encoder SESTrans.

\begin{figure*}[h]
    \centering
    \subfigure{

    \includegraphics[width=0.47\linewidth]{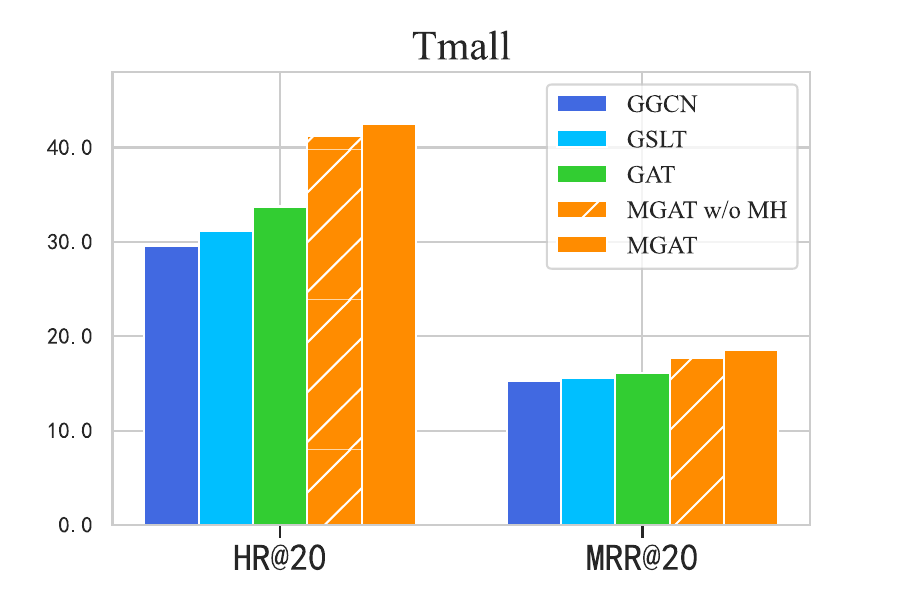}
    }
    \subfigure{

    \includegraphics[width=0.47\linewidth]{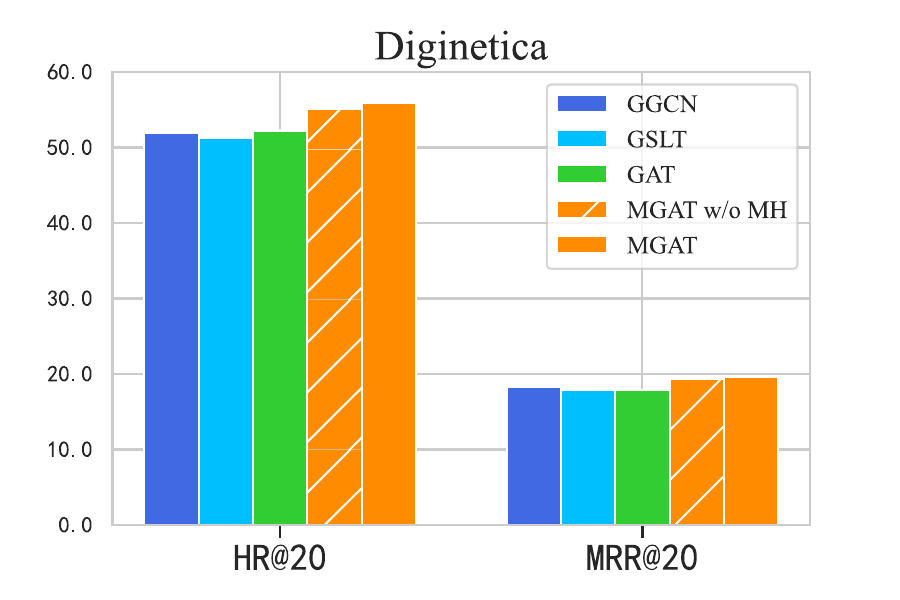}
    }
    \vspace{-0.4cm}

    \centering
    \subfigure{

    \includegraphics[width=0.47\linewidth]{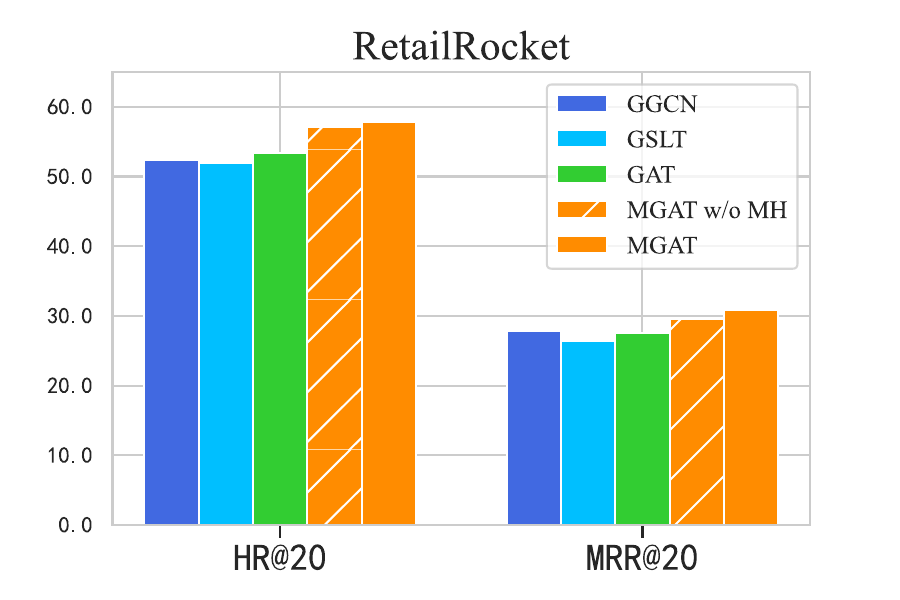}
    }
    \subfigure{
 
    \includegraphics[width=0.47\linewidth]{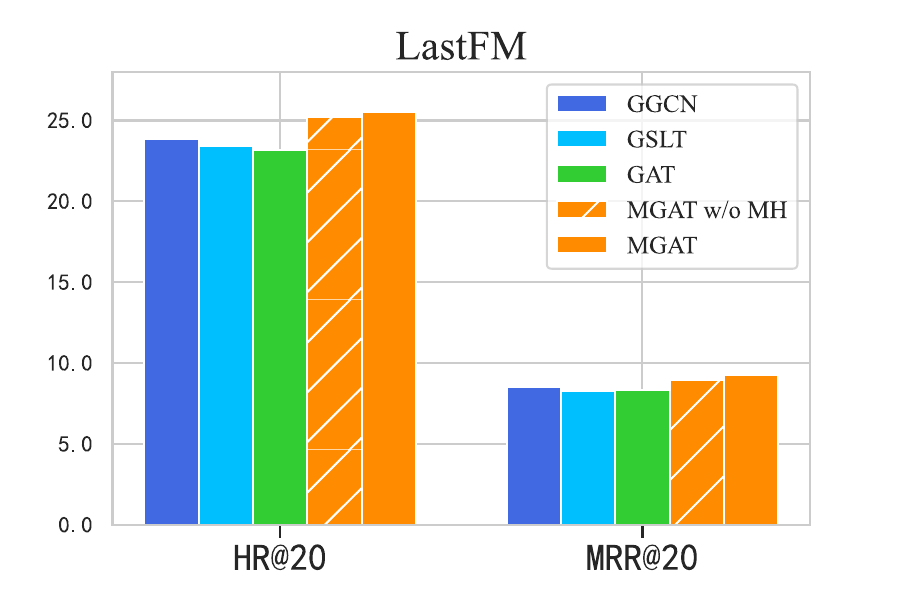}
    }
    \vspace{-0.4cm}
    
    \caption{Results of RESTC with different GNN-based spatial encoders. GSLT denotes GraphSAGE-LSTM, MGAT w/o MH denotes the single-head MGAT.}\label{fig:GNN_variant}
\end{figure*}

\subsubsection{\bruce{\textbf{Ablation on Different Spatial Encoder backbones (RQ3)}} }

Since our proposed RESTC is a model-agnostic framework that can effectively adapt to various GNN-based spatial encoders, we want to investigate the effectiveness of leveraging MGAT to learn the spatial representation of the session graph. Therefore, we compare it with other GNN-based backbones on Tmall, Diginectica, RetailRocket and LastFM. Specifically, we substitute MGAT backbone with some variants, including Graph Gate Neural Network (GGNN)~\cite{DBLP:conf/aaai/WuT0WXT19, DBLP:conf/ijcai/XuZLSXZFZ19}, GraphSAGE-LSTM~\cite{DBLP:conf/nips/HamiltonYL17}, GAT~\cite{DBLP:conf/iclr/VelickovicCCRLB18, DBLP:conf/cikm/QiuLHY19} and MGAT without multi-heads attention. Among them, GGCN constructs the session as a weighted directed graph and uses the occurrence frequency of item-pair transitions as edges and applies gate-based aggregate function; GraphSAGE-LSTM and GAT also adopt the same method to construct the session graph, but they utilize LSTM and attention weighted sum as the aggregation functions, respectively. As depicted in the Fig~\ref{fig:GNN_variant}, RESTC equipped with MGAT as \xin{a} spatial encoder is superior to all the comparative GNN-based backbones. Concretely, the MGAT backbone significantly improved compared with GAT and MGAT w/o MH, verifying the advantage of constructing sessions as multi-relational session graphs and leveraging multi-head MGAT. Moreover, compared to the GraphSAGE-LSTM and GGCN, MGAT achieves better performance, suggesting that the attention mechanism is more powerful for learning the spatial structural representation for the session graph.

\subsection{Further Analysis on Spatio-Temporal Contrastive Learning (RQ4)}

To further analyze what factors affect the performance of our proposed spatio-temporal contrastive learning, we move on to studying different settings. We first investigate the impact of temperature $\tau$. Then, we dive into the influence of distinct negative sampling strategies in the contrastive learning objective function. We adjust the hyperparameter $\tau$ on Tmall and Diginetica, which have a similar trend to other datasets. Then, we demonstrate the results of using variants of negative sampling on Tmall, Diginetica, RetailRocket, and LastFM due to the limited space.





\begin{figure*}[t]
    \centering
    \subfigure{
    \includegraphics[width=0.47\linewidth]{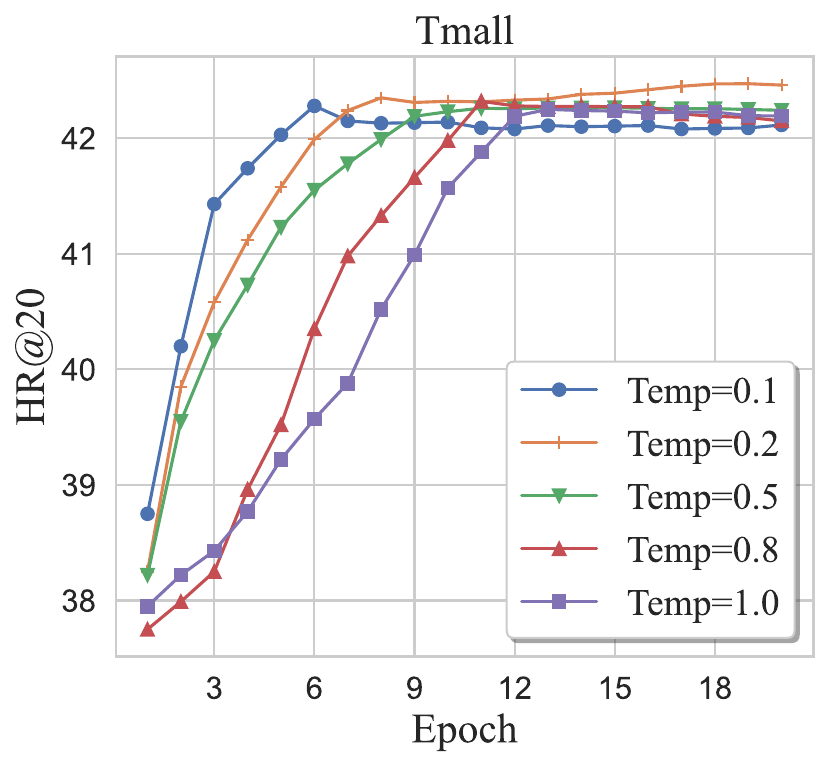}
    }
    \subfigure{
    \includegraphics[width=0.47\linewidth]{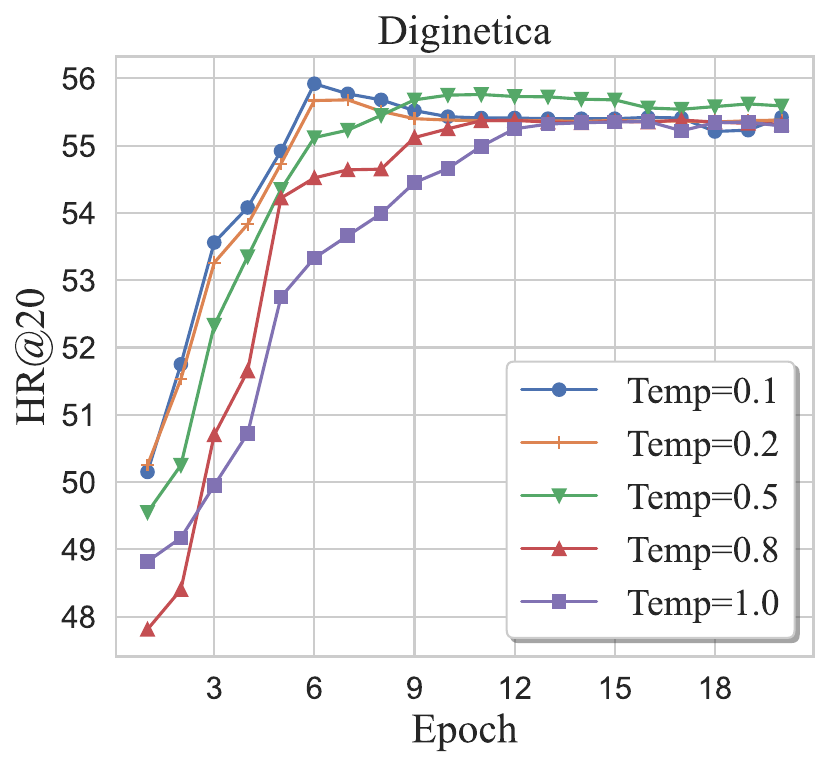}
    }
    \vspace{-0.4cm}
    
   \caption{Model performance of RESTC with different temperature $\tau$. }\label{fig:tau}
\end{figure*}

\subsubsection{Impact of Temperature $\tau$}

As mentioned in~\cite{DBLP:conf/icml/ChenK0H20, DBLP:conf/sigir/WuWF0CLX21}, $\tau$ plays a critical role in hard negative mining for contrastive learning. The experiment results in Fig~\ref{fig:tau} show the curves of RESTC performance with respect to different $\tau$. We can observe that: (1) The larger the value of $\tau$ (e.g., 1.0), the slower the model converges during training, and there is a significant decrease for the model's performance when it converges. Similar to~\cite{DBLP:conf/sigir/WuWF0CLX21}, we attribute this phenomenon to the difficulty of identifying hard negative samples, whose temporal representations are similar to that of positive samples, thus making the model fail to distinguish them from the positive samples in the latent space. (2) In contrast, adjusting $\tau$ with a too small value (e.g., 0.1) will cause the model to converge quickly, leading to prematurely overfitting during training. We conjecture the small $\tau$ could make the model focus excessively on the hard negative samples and offer more gradients to guide the optimization, thus making the spatial and temporal representations easier to discriminate then accelerate the training process~\cite{DBLP:conf/wsdm/RendleF14}. Therefore, depending on the dataset, we choose the value of $\tau$ between 0.1 and 1.




\subsubsection{Variants of Negative Sampling Strategy}

\begin{figure*}[htbp]
  \centering
  \includegraphics[width=1\linewidth]{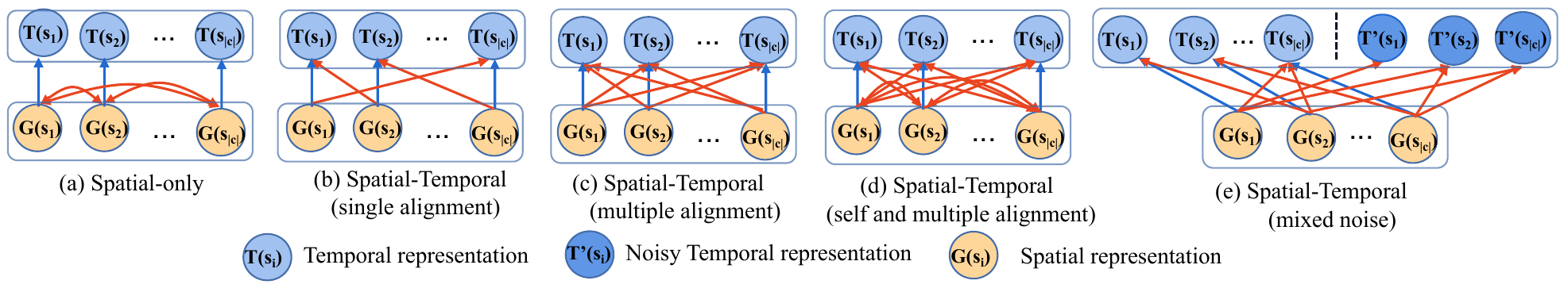}
  \vspace{-0.2in}
  \caption{Four variants of negative sampling strategy and the default method.}
  \vspace{-0.2in}
  \label{fig:neg_sam}
\end{figure*}

To investigate how the choices of negative sampling affect the performance of contrastive learning, we ablate on several negative sampling strategies as shown in Fig~\ref{fig:neg_sam}. Specifically, we compare our default method with four variants of session-level contrastive learning, which select negative samples from spatial or temporal session representations in a training batch: (a) \textit{Spatial-only}, which selects the representations of other sessions in the spatial candidates \textit{Spatial-only}; (b) \textit{Spatio-Temporal (single alignment)}, which randomly selects one different temporal presentation, denoted as S-T (sa) ; (c) \textit{Spatio-Temporal (multiple alignments)}, which selects the other temporal representations from the batch, denoted as S-T (ma); (d) \textit{Spatio-Temporal (self and multiple alignments)}, which selects the representations of both spatial and temporal candidates in the batch, denoted as S-T (sma). As illustrated in Sec~\ref{sec:contrastive}, the default method of RESTC is \textit{Spatio-Temporal (mixed noise)}.

From Table~\ref{tab:negative_sampling}, we can observe that the Spatial-only method performs worse than all the comparative methods, which only use spatial representations as negative samples. This may indicate that without using temporal representations as negative samples, it will be challenging to align spatio-temporal information in the latent space, leading to sub-optimal performance. Besides, S-T (ma) and S-T (sma) slightly perform better than S-T (sa), which we conjecture is because, increasing the sampling size and diversity of negative samples (spatial and temporal views) facilitates the model to distinguish between positive and negative sample pairs. In addition, S-T (mn) outperforms all the variants of sampling strategies, which may be because adding random noise to the set of temporal representations is beneficial to enhance the robustness of contrastive learning. Moreover, we also validate the correlation between noise sampling strategy and bach size. The results are consistent with SimCLR~\cite{DBLP:conf/kdd/ChenW20}, enlarging the number of negative samples by increasing the batch size from 128 to 512 significantly improves performance.

\begin{table}[htbp]
\caption{Comparison on Variants of Negative Sampling.}
\vspace{-0.1in}
 \scalebox{0.75}{
  \begin{tabular}{l|cc|cc|cc}
\midrule[1.2pt]
Dataset & \multicolumn{2}{|c|}{ \textbf{TM} } & \multicolumn{2}{c|}{ \textbf{DG} } & \multicolumn{2}{c}{ \textbf{RR} } \\
\midrule
Measures & \multicolumn{2}{|c|}{ HR@20 MRR@20 } & \multicolumn{2}{c|}{ HR@20 MRR@20 } & \multicolumn{2}{c}{ HR@20 MRR @20 } \\
\midrule
 Spatial-only & $41.33$ & $17.96$ & $54.95$ & $19.39$ & $57.19$ & $30.15$ \\
 S-T (sa)    & $41.95$ & $18.11$ & $55.33$ & $19.44$ & $57.35$ & $30.19$ \\
 S-T (ma)   & $42.23$ & $18.18$ & $55.56$ & $19.42$ & $57.45$ & $30.52$ \\
 S-T (sma)   & $42.35$ & $18.23$ & $55.52$ & $19.45$ & $57.38$ & $30.44$ \\
 \midrule
 
 S-T (mn)(bz=128) & $41.95$ & $18.15$ & $55.32$ & $19.44$ & $57.23$ & $30.32$ \\
 S-T (mn)(bz=256) & $42.22$ & $18.21$ & $55.56$ & $19.53$ & $57.41$ & $30.56$ \\
 S-T (mn)(bz=512) & $\textbf{42.47}$ & $\textbf{18.52}$ & $\textbf{55.93}$ & $\textbf{19.65}$ & $\textbf{57.81}$ & $\textbf{30.82}$ \\
 
\midrule[1.2pt]
\end{tabular}}\label{tab:negative_sampling}

\begin{tablenotes}
        \setlength\labelsep{0pt}
	\begin{footnotesize}
	\item
         sa, ma, sma and mn denote single alignment, multiple alignment, self and multiple alignment and mixed noise, respectively.
        \par
	\end{footnotesize}
      \end{tablenotes}
\vspace{-0.2in}
\end{table} 

\subsection{Impact of Hyperparameters (RQ6)} \label{sec: hyparameter_analysis}

\begin{figure*}[h]
    \centering
    \subfigure{
    \includegraphics[width=0.47\linewidth]{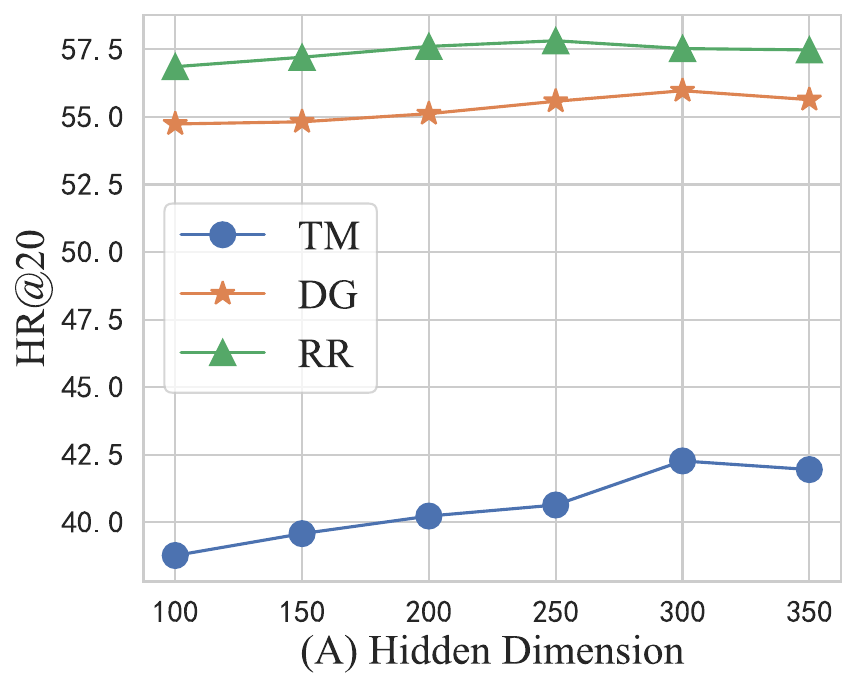}
    }
    \subfigure{
    \includegraphics[width=0.47\linewidth]{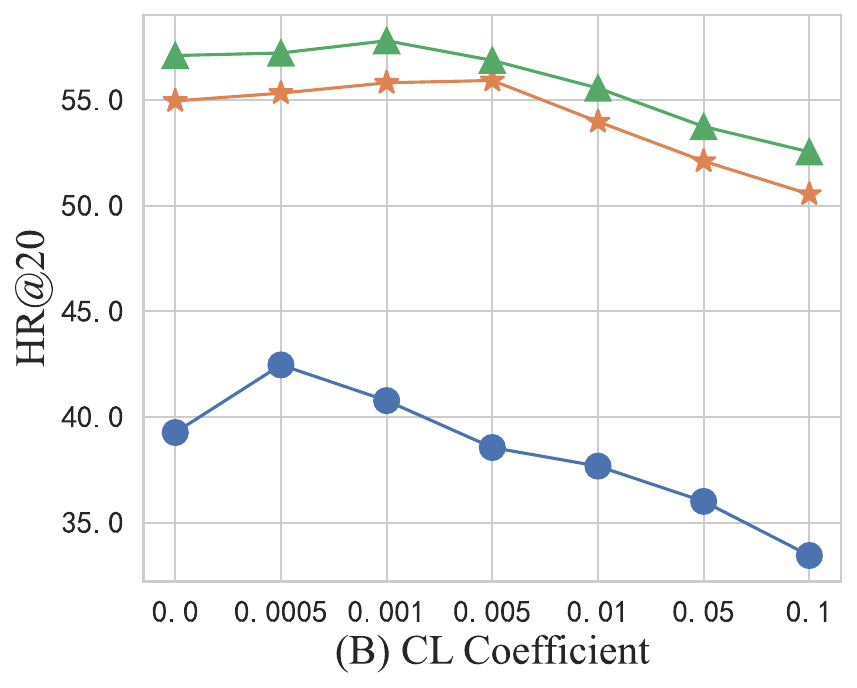}
    }
    \vspace{-0.4cm}
    
    \subfigure{
    \includegraphics[width=0.47\linewidth]{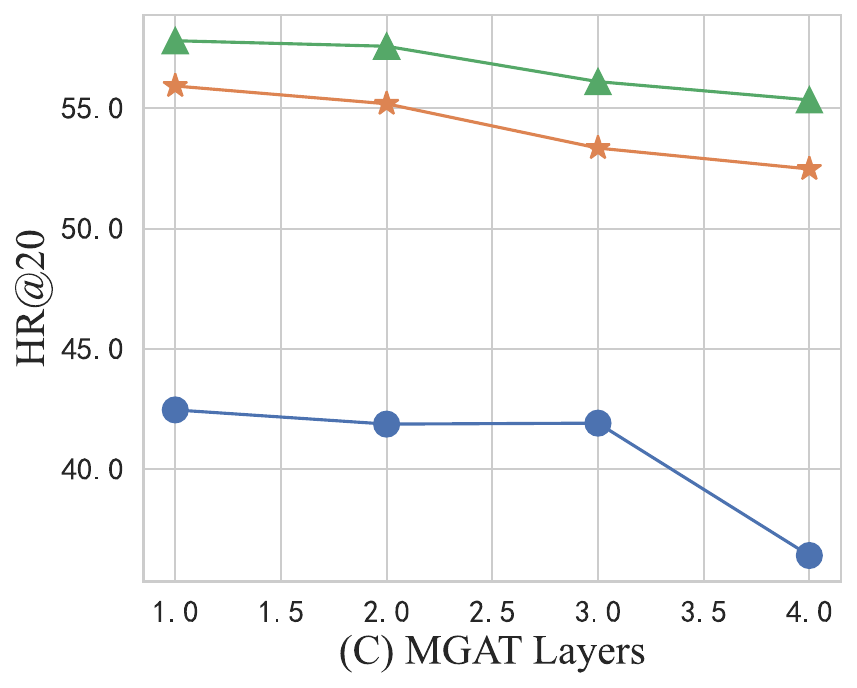}
    }
    \subfigure{
    \includegraphics[width=0.47\linewidth]{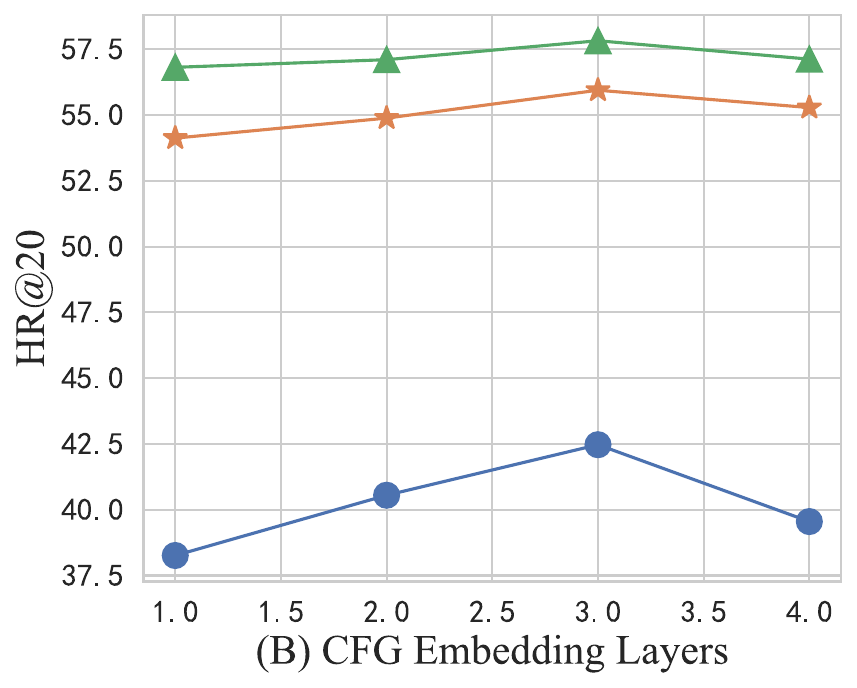}
    }
    \vspace{-0.4cm}
    
   \caption{Hyperparameter analysis of RESTC.}\label{fig:hyperparameter1}
\end{figure*}

Next, we analyze the sensitivity of RESTC with different hyperparameter settings. Due to the limited space, we only show the result of HR@20 on Tmall, Diginetica, and Retailrocket.

\begin{figure*}[h]
    \centering
    \subfigure{
   
    \includegraphics[width=0.47\linewidth]{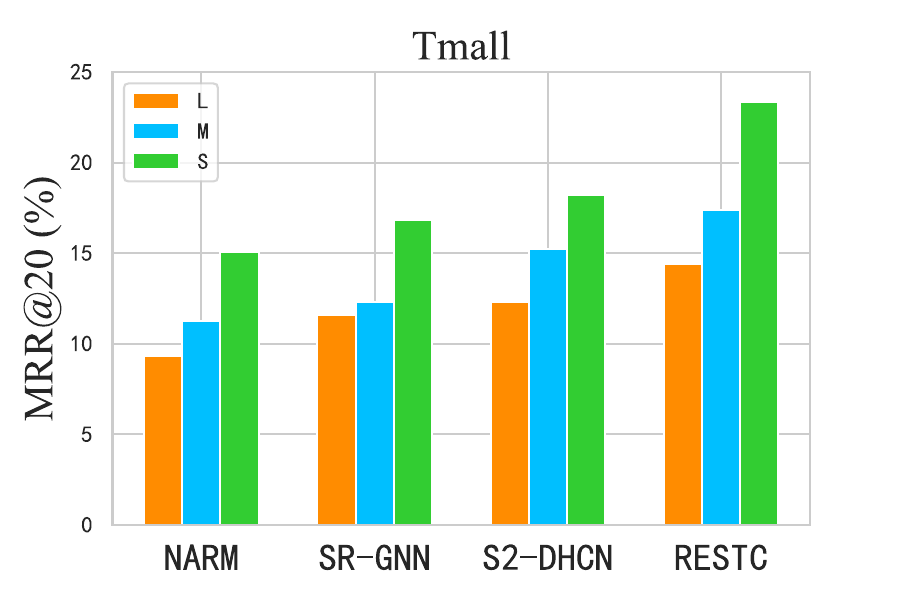}
    }
    \subfigure{

    \includegraphics[width=0.47\linewidth]{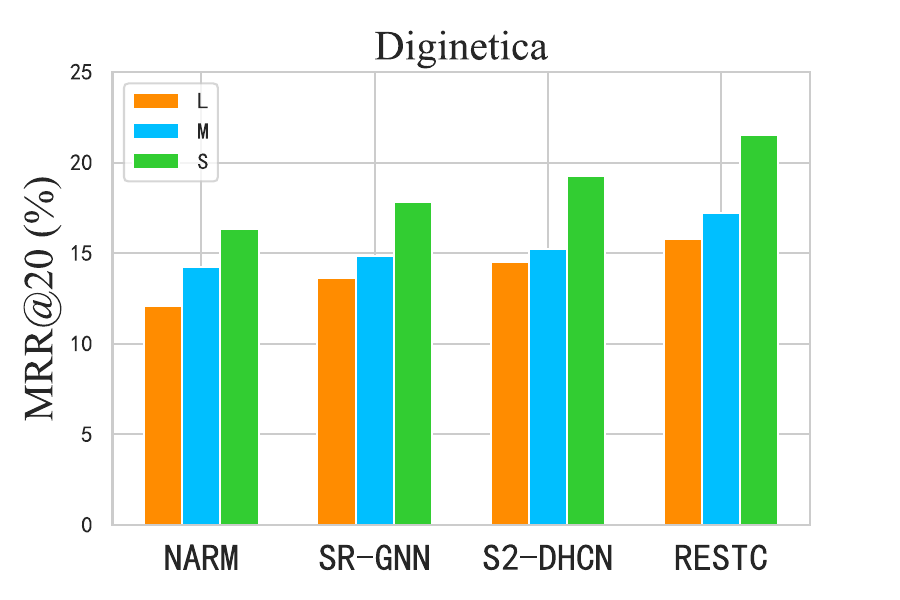}
    }
    \vspace{-0.4cm}
    
    \centering
    \subfigure{

    \includegraphics[width=0.47\linewidth]{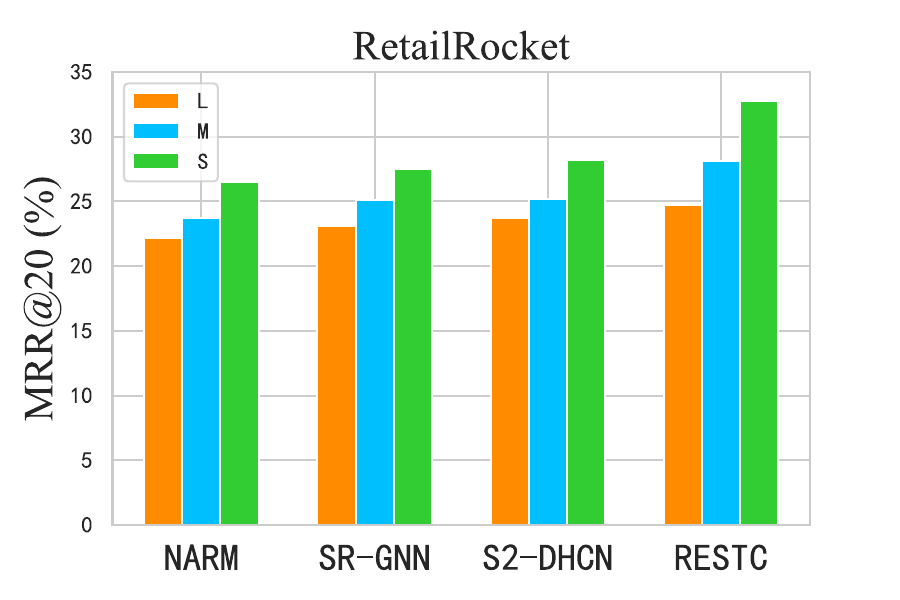}
    }
    \subfigure{
 
    \includegraphics[width=0.47\linewidth]{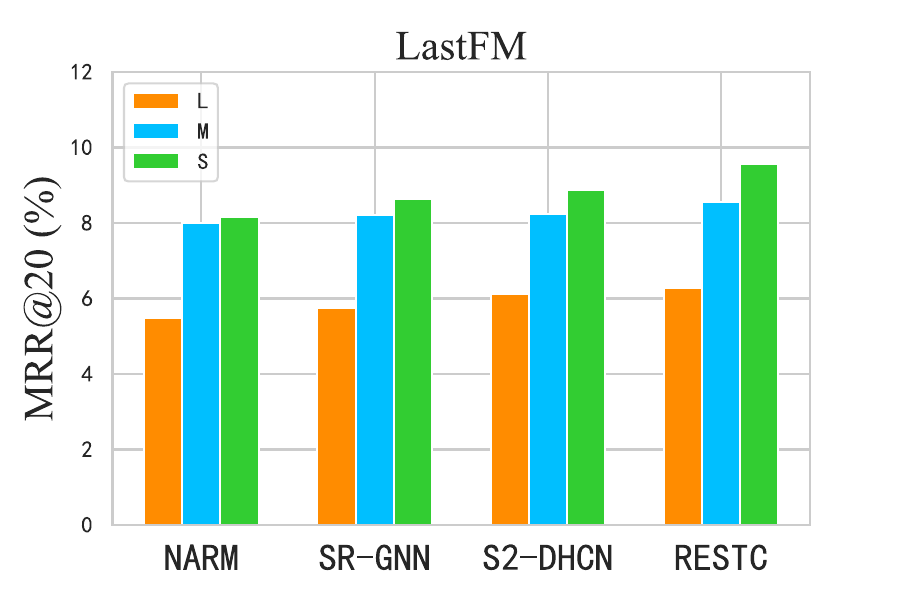}
    }
    \vspace{-0.4cm}
    
    \caption{MRR@20 on sessions of different lengths. }\label{fig:mrr_length}
\end{figure*}

\subsubsection{Impact of Hidden Dimension} 

\noindent To investigate the impact of hidden dimension, we test the performance when increasing the size from 100 to 400. From the leftmost of Fig.~\ref{fig:hyperparameter1} (A), we can conclude that increasing the hidden dimension does not continuously improve the performance. Our RESTC model achieves the best performance in 300 for Diginetica and Tmall while obtaining an optimal result in 200 for Retailrocket. The reason might be that a larger hidden size might lead to overfitting.

\subsubsection{Strength of Contrastive Learning}

\noindent In RESTC, we utilize the hyperparameter $\eta_{1}$ to trade off the contrastive loss and the cross entropy loss. To demonstrate the utility of $\eta_{1}$, we compare the experimental results by using the $\eta_{1}$ values from $ [ 0.0, 0.0005, 0.001, 0.005, $ $0.01, 0.05, 0.1 ]$. As in the rightmost of Fig~\ref{fig:hyperparameter1} (B), larger $\eta_{1}$ does not show a tendency of better performance. Our model obtains the most satisfactory performance when $\eta_{1}$ is near 0.005, 0.001, 0.0005 for Diginetia, Retailrocket, and Tmall, respectively. The  HR@20 drops obviously when the $\eta_{1}$ becomes larger than these values, especially in Tmall. The main reason is that increasing $\eta_{1}$ might harm the optimization of the main prediction task. Therefore, according to grid search, we set the corresponding coefficient $\eta_{1}$.

\subsubsection{Effect of MGAT \xin{Layers}}
\wan{\noindent To further analyze the impact of the aggregation layer numbers of the spatial encoder MGAT, we vary the number of MGAT layers in the range of $\left\{1, 2, 3, 4\right\}$. As the results presents in Fig~\ref{fig:hyperparameter1} (C), leveraging 1 layer MGAT for RESTC has already achieved the best performance, and stacking more layers leads to \xin{a} decreasing tendency. We conjecture that adopting more layers will cause \xin{the overfitting} issue since most of the sessions are relatively \xin{shorter} according to the average lengths of the dataset statistics in Table~\ref{tab:data}.}
 
\subsubsection{Effect of CFG Embedding \xin{Layers}}

\noindent The embedding of a Collaborative Filtering Graph (CFG) enriches the current session with inter-session information, which is an efficient way to solve data sparsity problems and enhance recommendation performance. We range the layer numbers from 1 to 4 to study the impacts of the CFG embedding module's depth. From the middle of Fig.~\ref{fig:hyperparameter1}, we observe that the three-layer setting makes RESTC obtain the best result. And stacking more layers will add more noise information to the over-smoothing issue of high-order relations of graphs.




\begin{figure*}[h]
    \centering
    \subfigure{
   
    \includegraphics[width=0.47\linewidth]{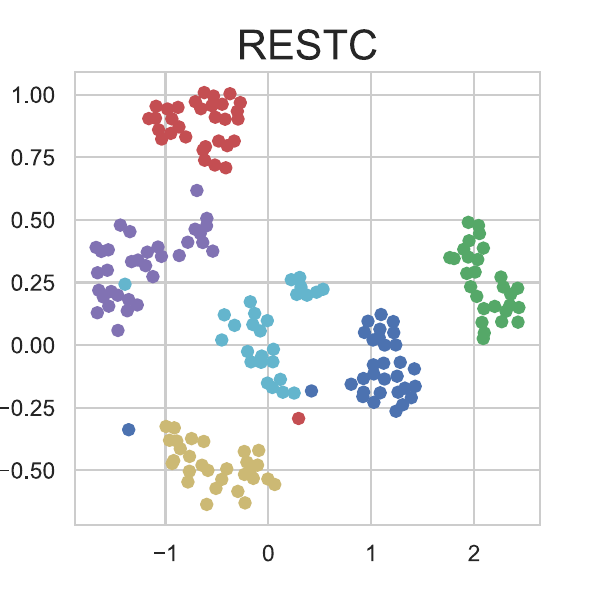}
    }
    \subfigure{

    \includegraphics[width=0.47\linewidth]{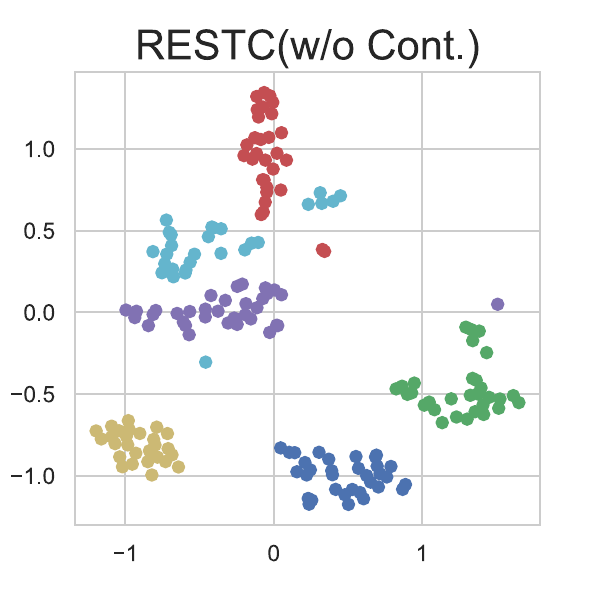}
    }
    \vspace{-0.4cm}
    
    \centering
    \subfigure{

    \includegraphics[width=0.47\linewidth]{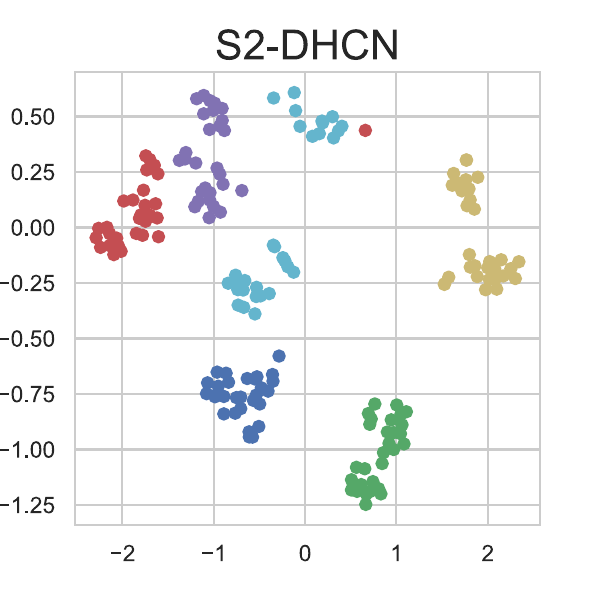}
    }
    \subfigure{
 
    \includegraphics[width=0.47\linewidth]{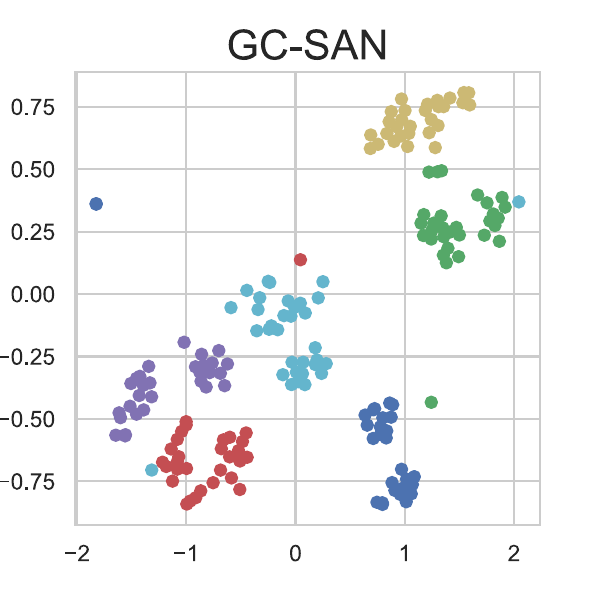}
    }
    \vspace{-0.4cm}
    
    \caption{t-SNE visualization of session embedding in a latent
space, each color represents a specific label.}\label{fig:tsne}
\end{figure*}

\subsection{Analysis on Different Session Lengths (RQ5)}

In many scenarios, sessions are transferred to the server at various lengths~\cite{DBLP:journals/tois/PanCCC22}. It is worthwhile to investigate the robustness of our RESTC model compared with baselines on different lengths of sessions. 
We separate all the sessions in Tmall, Diginetica, RetailRocket and LastFM into three groups, \textbf{short group} (S) with length of sessions from 0 to 5, \textbf{medium group} (M) with sessions from 5 to 10, rest of sessions are in the \textbf{ long group} (L). We utilize MRR@20 to evaluate the performance of the methods instead of HR@20 since the MRR metric can better reflect the ranking quality of correct results. Fig.~\ref{fig:mrr_length} demonstrates that RESTC outperforms \wan{the selective sequence-based baseline NARM, the GNN-based baseline SR-GNN, and the contrastive learning augmented GNN baseline S$^2$-DHCN with different lengths of sessions.} Note that all methods have performance drops when session length increases. This may be because long item transitions are difficult to model users' preferences since the diversity of users' intents or missed clicks in the long sequence. The results also indicate the superiority of RESTC in scenarios when the ongoing session is short because of its effectiveness in handling data sparsity with CFG embedding in Sec.~\ref{sec:cfg_embedding}.

\subsection{Representation Quality of RESTC (RQ7)}
To evaluate whether spatio-temporal contrastive learning affects the representation learning performance, we utilize t-SNE to reduce the dimension of learned embeddings and visualize them in 2D planes. As shown in Fig.~\ref{fig:tsne}, we compare the visualize results of RESTC, RESTC(w/o Cont.), S$^2$-DHCN and GC-SAN on Retailrocket and leverage six labels and randomly sample 50 session instances for each label. It is expected that session embeddings should be closer if they have the same label (next-to-click item). \wan{From Fig.~\ref{fig:tsne}, by comparing RESTC and its variant, we observe that removing spatio-temporal contrastive learning makes the learned embedding more indistinguishable in the latent space, showing that contrastive learning makes a better alignment for RESTC between session embeddings w.r.t. the same label. Moreover, some session embeddings with different labels are mixed to some degree for S$^2$-DHCN and GC-SAN, which \xin{makes} them indiscernible. In contrast, our RESTC shows a more diverse distribution and hence can better make a correct prediction, demonstrating the superiority of RESTC in better representation learning.}

\section{Conclusion}
This paper proposes a novel framework called RESTC, which aims to effectively learn the session representation from cross-view \xin{interactions} and collaborative filtering information. It is equipped with spatio-temporal contrastive learning to extract self-supervised signals from spatial and temporal views to mitigate temporal information loss and improve the quality of representation learning. In the next-item prediction task, we utilized the embedding of the collaborative filtering graph to enrich the spatial structure information, which can also solve the data sparsity problem of the short-term session. Extensive experiment results demonstrate that RESTC achieves significant improvements compared with other recent baselines.


\bibliographystyle{ACM-Reference-Format}
\bibliography{ref}

\end{document}